\documentclass[fleqn,usenatbib]{mnras}

\usepackage{newtxtext,newtxmath}

\usepackage[T1]{fontenc}

\DeclareRobustCommand{\VAN}[3]{#2}
\let\VANthebibliography\thebibliography
\def\thebibliography{\DeclareRobustCommand{\VAN}[3]{##3}\VANthebibliography}

\usepackage{graphicx}	
\usepackage{amsmath}	

\newcommand{\gadgetthree}{\mbox{GADGET-3}}
\newcommand{\gadgetfour}{\mbox{GADGET-4}}
\newcommand{\ketju}{KETJU}
\newcommand{\codewebsite}{\url{https://www.mv.helsinki.fi/home/phjohans/ketju}}

\newcommand{\pn}[1]{#1PN}
\newcommand{\Msun}{\ensuremath{\mathrm{M}_{\sun}}}

\newcommand{\rreg}{\ensuremath{r_\mathrm{region}}}

\newcommand{\dd}[1]{\ensuremath{\mathrm{d}#1}}
\newcommand{\dv}[2]{\ensuremath{\frac{\dd{#1}}{\dd{#2}}}}
\newcommand{\vb}[1]{\ensuremath{\mathbfit{#1}}}
\newcommand{\bigO}{\mathcal{O}}

\graphicspath{{figures/}}

\title[KETJU for SMBH dynamics in GADGET-4]{KETJU -- resolving small-scale supermassive black hole dynamics in GADGET-4}

\author[M. Mannerkoski et al.]{%
Matias Mannerkoski,$^{1}$\thanks{E-mail: matias.mannerkoski@helsinki.fi}
Alexander Rawlings,$^{1}$
Peter H. Johansson,$^{1}$\thanks{E-mail: peter.johansson@helsinki.fi}
Thorsten Naab,$^{2}$
Antti Rantala,$^{2}$
\newauthor%
Volker Springel,$^{2}$
Dimitrios Irodotou,$^{1}$
Shihong Liao$^{1}$
\vspace*{0.1cm}\\%
$^{1}$
Department of Physics, University of Helsinki, Gustaf H\"allstr\"omin katu 2, FI-00014 Helsinki, Finland
\\%
$^{2}$
Max-Planck-Institut f\"ur Astrophysik, Karl-Schwarzchild-Str 1,
D-85748 Garching, Germany
}

\date{Accepted XXX. Received YYY; in original form ZZZ}

\pubyear{2023}

\begin{document}
\label{firstpage}
\pagerange{\pageref{firstpage}--\pageref{lastpage}}
\maketitle

\begin{abstract}
We present the new public version of the \ketju{} supermassive black hole (SMBH) dynamics module, as implemented into \gadgetfour{}.
\ketju{} adds a small region around each SMBH where the dynamics
of the SMBHs and stellar particles are integrated using an algorithmically
regularised integrator instead of the leapfrog integrator with gravitational softening used by \gadgetfour{}.
This enables modelling SMBHs as point particles even during close interactions with stellar particles or other SMBHs,
effectively removing the spatial resolution limitation caused by gravitational softening.
\ketju{} also includes post-Newtonian corrections, which allows following the dynamics of SMBH binaries to
sub-parsec scales and down to tens of Schwarzschild radii.
Systems with multiple SMBHs are also supported, with the code also including the leading non-linear cross terms
that appear in the post-Newtonian equations for such systems.
We present tests of the code showing that it correctly captures,  at sufficient mass resolution, the sinking driven by dynamical friction
and binary hardening driven by stellar scattering.
We also present an example application demonstrating how the code can be applied to study the dynamics
of SMBHs in mergers of multiple galaxies and the effect they have on the properties of the surrounding galaxy.
We expect that the presented \ketju{} SMBH dynamics module can also be straightforwardly incorporated into other codes similar to \gadgetfour{},
which would allow coupling small-scale SMBH dynamics to the rich variety of galactic physics models that exist in the literature.
\end{abstract}




\section{Introduction}

Supermassive black holes (SMBHs) with masses in the range $M_{\bullet}=10^{6}\textnormal{--}10^{10} \, \Msun$ are found
in the centres of all massive galaxies in the local Universe (e.g. \citealt{2005SSRv..116..523F,kormendy2013}).
In the $\Lambda$CDM model, galaxies grow through mergers and gas accretion \citep{1978MNRAS.183..341W,2017ARA&A..55...59N}, and
this hierarchical growth will thus invariably result in situations with multiple SMBHs in the same galaxy
(e.g. \citealt{2003ApJ...582..559V}).

The SMBH merger process proceeds through three main phases \citep{begelman1980,2013degn.book.....M}.
First, the SMBHs will sink to the centre of the merger remnant due to dynamical friction and will form a bound
binary with a semimajor axis of some tens of parsecs, depending on the masses of the SMBHs. In the second phase the SMBH binary will experience complex
three-body interactions with individual stars that drain energy and angular momentum away from the binary \citep{1980AJ.....85.1281H}.
At this stage the additional drag from gas in the form of a circumbinary disc is also expected to impact the evolution of the binary
\citep[e.g.][]{2014ApJ...783..134F,2020ApJ...901...25D}. In the final third stage the SMBH binary will be driven to coalescense by gravitational wave (GW) emission
at sub-parsec separations \citep{1963PhRv..131..435P}.

Although the basic dynamical framework for SMBH merging is rather well understood, there remain several challenges. For example, one of the challenges is to
numerically resolve the detailed interactions of the SMBHs and the stellar component that result in SMBH binary scouring and the formation
of galactic cores \citep[e.g.][]{2018ApJ...864..113R,2021MNRAS.502.4794N}. Modelling the evolution of the SMBH binaries is especially challenging in simulations that simultaneously
include a gaseous component
\citep[e.g.][]{2013MNRAS.429.3114C,2015MNRAS.449..494R,2015MNRAS.447.2123C},
with one option being that one resorts
to an approach that involves the use of on-the-fly code switching \citep{2016ApJ...828...73K,2018ApJ...868...97K}.
Another challenge is to provide detailed predictions for the GWs emitted by SMBH binaries
that could be observed by instruments such as the ground-based Pulsar Timing Arrays (PTAs, e.g.\ \citealp{2018ApJ...859...47A}) and the future
Laser Interferometer Space Antenna (LISA, \citealp{2023LRR....26....2A}).

The SMBH merger process has been traditionally simulated using direct summation $N$-body codes that are well suited for studying
collisional stellar systems with typically $N\lesssim 10^{6}$ particles
\citep[e.g.][]{1991Natur.354..212E,2001ApJ...563...34M,2006ApJ...642L..21B,2008ApJ...678..780G,2011ApJ...732...89K,2015ApJ...810..139H}.
The codes employed for these
type of simulations typically use sophisticated high-order integration schemes, including in some instances also regularisation, which
allows low or zero softening for the gravitational interactions involving SMBHs, and thus enables following the evolution of SMBH binaries
to very small spatial scales  \citep[e.g.][]{2009ApJ...695..455B,2015ApJ...810...49V,2017MNRAS.464.2301G,2022MNRAS.511.4753G}.
However, these types of pure $N$-body codes do not generally include the capability to simulate the dynamics of gas and the associated astrophysical processes, thus limiting the applicability of these codes for a self-consistent treatment of SMBH dynamics in a full galactic environment.
Another fundamental limitation of many direct $N$-body codes is the number of particles, which is typically limited to a few million for most applications,
due to the steep scaling of the computational time with particle number, $\bigO(N^2)$. However, this limit is steadily, but
slowly, rising with improvements both in the employed software and the available computational resources
\citep[e.g.][]{1999PASP..111.1333A,2008MNRAS.389....2H,2014ComAC...1....1D,2015MNRAS.450.4070W,2020MNRAS.497..536W,2021MNRAS.502.5546R}.

In recent years there has also been significant progress in simulating the evolution and impact of SMBHs in both galaxy mergers \citep[e.g.][]{2005MNRAS.361..776S,2009ApJ...690..802J,2009ApJ...707L.184J,2012ApJ...754..125C} and in a
full cosmological setting \citep[e.g.][]{2007MNRAS.380..877S,2009MNRAS.398...53B,2014MNRAS.442.2304H,2015MNRAS.452..575S} using initially smoothed particle hydrodynamics (SPH, \citealp{2005MNRAS.364.1105S,2005MNRAS.361..776S}) and also later adaptive mesh refinement \citep{2002A&A...385..337T,2012MNRAS.420.2662D}
and moving mesh codes \citep{2010MNRAS.401..791S,2013MNRAS.436.3031V}. These simulations allow for a very large number of particles, as the gravitational forces are typically resolved with either a tree
or a mesh resulting in a typical scaling of $\bigO(N \log N)$ of the computational time. In addition, these simulation codes are able to model
astrophysical processes by including sophisticated subresolution models for gas physics, star formation and the feedback from evolving stellar populations and
SMBHs.

However, a fundamental limitation of this approach is the necessary inclusion of a gravitational softening or equivalently a minimum grid cell that sets a natural resolution limit, below which the dynamics cannot be accurately resolved. A possible solution to circumvent the effects of softening is to add a
subresolution drag force to the equations of motion to account for the unresolved dynamical friction \citep{2015MNRAS.451.1868T,2022MNRAS.510..531C,2023MNRAS.519.5543M}. However, in most applications
the merging of SMBHs below the softened resolution limit has instead been modelled using semi-analytic methods
\citep[e.g.][]{2017MNRAS.464.3131K,2017MNRAS.471.4508K,2019MNRAS.486.4044B,2022MNRAS.509.3488I}, which have to
make several assumptions about the unresolved binary orbits.

The hybrid method is an alternative approach that combines the best aspects of direct summation codes and softened galaxy formation codes by linking together the
lower-accuracy hydrodynamical galaxy formation codes used for cosmological simulations,
such as the various versions of GADGET and related codes, with a more accurate integration scheme used only in the vicinity of SMBHs.
This allows for simultaneously modelling global galactic-scale dynamical and astrophysical processes, while solving the dynamics of SMBHs and the surrounding
stellar systems at sub-parsec resolution. In addition, a hybrid code enables running isolated galaxy mergers without gas at higher resolution and in
greater number than is possible with the more traditional types of direct $N$-body codes. Several such hybrid codes have been developed, for example
\citet{1989ApJS...71..871J} and \citet{1993ApJ...414..200M} combined a tree code with a regularisation algorithm, whereas \citet{2011PASJ...63..881O} and \citet{2015ComAC...2....6I} also combined a tree algorithm with a direct summation
code, but without the inclusion of regularisation. There are also code frameworks, such as BRIDGE \citep{2007PASJ...59.1095F} and AMUSE (The Astrophysical Multipurpose Software Environment, \citealp{2013A&A...557A..84P}) which enable the combination of different types of $N$-body codes for the same simulation problems. For \ketju{} the most direct precursor code is the rVINE code
\citep{2015MNRAS.452.2337K}, which combines algorithmically regularised integration with the VINE code, which is a tree/SPH code employing a binary tree algoritm \citep{2009ApJS..184..298W}.

The first version of \ketju{} \citep{2017ApJ...840...53R} was implemented within the \gadgetthree{} code. This version of the code was used
to run isolated collisionless galaxy mergers in order to study the formation of cores in massive galaxies \citep{2018ApJ...864..113R,2019ApJ...872L..17R} and for calculating the GW signal from inspiraling SMBHs in galactic-scale simulations \citep{2019ApJ...887...35M}. The code was subsequently improved, with the main updates
including the replacement of the original regularised AR-CHAIN integrator \citep{2008AJ....135.2398M} with the MSTAR integrator \citep{2020MNRAS.492.4131R}, which resulted in a significant performance
improvement. In addition, the code interface was updated, especially as regarding the treatment of hydrodynamics and feedback, thus enabling both cosmological
zoom-in simulations with tens of SMBHs \citep{2021ApJ...912L..20M,2022ApJ...929..167M} and isolated merger simulations of gas-rich galaxies \citep{2023MNRAS.520.4463L}, including now also the effects of gas cooling, star
formation, stellar feedback and crucially the feedback from SMBH binary systems \citep{2023MNRAS.520.4463L}.

While the original version of \ketju{} was implemented in \gadgetthree{},
in this paper we instead discuss a new version of \ketju{}, as implemented in \gadgetfour{}.
Unlike \gadgetthree{} which was never publicly released, \gadgetfour{} \citep{2021MNRAS.506.2871S}
was recently made publicly available. We have thus now implemented the \ketju{} SMBH dynamics module into \gadgetfour{} and together with this paper we make this
version of \ketju{} publicly available.\footnote{\codewebsite{}}.
In this paper we describe this new public implementation of \ketju{} and present several tests of code correctness
and performance, as well as an example application demonstrating how the \ketju{} code can be used to simulate complex systems with multiple merging SMBHs.

This paper is structured as follows. In Section~\ref{sec:Code_description} we describe the main features of the \ketju{} code, \gadgetfour{} and the regularised
MSTAR integrator, with a main emphasis on how the interface between the codes operates. In this Section we also present a schematic of how the \ketju{}
integration is performed. In Section~\ref{sec:Code_tests}, we perform various code tests by first demonstrating that \ketju{} correctly captures
dynamical friction and produces converged results in the hardening rate of SMBH binaries. This is followed by integrator tolerance and computational scaling tests.
Then, in Section~\ref{Sec:Formation_cores} as a \ketju{} demonstration we perform simulations of multiple merging galaxies containing SMBHs, in order to study the formation of extended galactic cores. Finally, in Section~\ref{sec:Conclusions} we present our conclusions.

\section{Code Description}
\label{sec:Code_description}

\subsection{\ketju{} overview}

\begin{figure}
\includegraphics[width=\columnwidth]{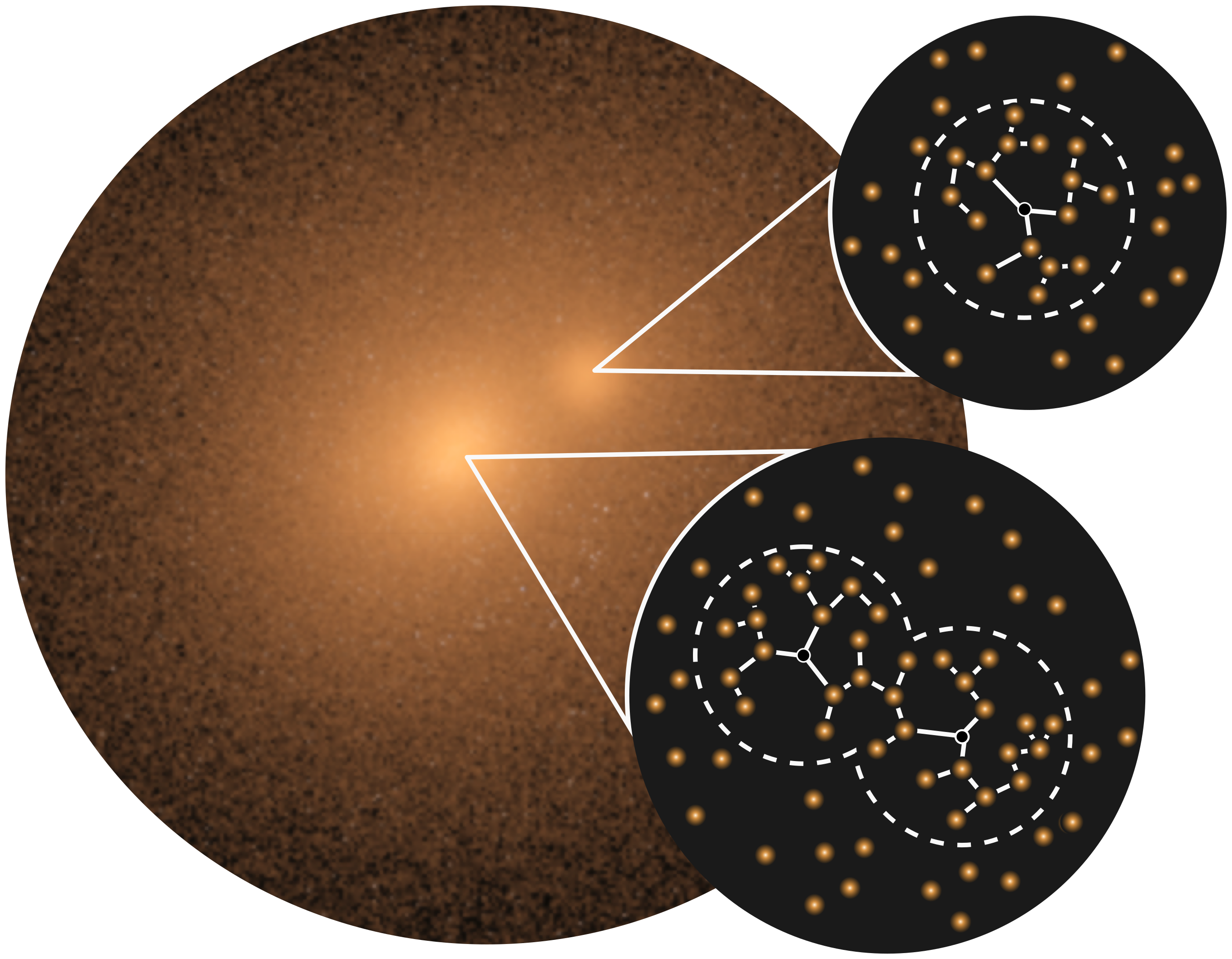}
\caption{
An illustration of \ketju{} applied to a system of three merging elliptical galaxies, where the first merger has already completed
resulting in an SMBH binary, with the third galaxy currently merging into the system.
The main idea in \ketju{} is to add small spherical regions (dashed circles) with typical radii of $\sim 10\,\mathrm{pc}$ centred on the SMBHs (shown as black dots
in the two insets),
where the dynamics are integrated using a high-accuracy regularised integrator.
This allows treating SMBHs as point particles even in close interactions with other particles,
allowing the small-scale dynamics to be modelled below the gravitational softening length used in other
parts of the simulation.
The surrounding kiloparsec-scale structure of the galaxies, shown in the background image,
is simulated using the standard \gadgetfour{} methods with softened gravity.
The regularised integrator uses a minimum spanning tree--based relative coordinate system
to reduce rounding errors in close encounters, indicated with the white lines between the particles.
The figure is reproduced from \citet{matias_thesis}.
}
\label{fig:ketju_illustration}
\end{figure}

The main purpose of \ketju{} is to allow capturing of the small-scale dynamics of SMBHs in large-scale galactic merger and cosmological simulations run with
galaxy formation codes, such as GADGET, by introducing a separate higher-accuracy integrator that is used to solve the dynamics in small regions around SMBHs.
The integrator used in \ketju{} utilises algorithmic regularisation \citep{1999CeMDA..74..287M,1999MNRAS.310..745M},
which enables the integration of close encounters between point masses without running into issues caused by the divergence of the Newtonian gravitational acceleration.
This allows modelling gravitational interactions with SMBHs using non-softened gravity,
thus overcoming the resolution limitation caused by gravitational softening that
is required by the methods used in codes such as GADGET.
In addition, the integrator includes post-Newtonian corrections for the interactions between SMBHs,
which allows it to directly model the SMBH merger process from the start of a galactic merger down to the few final orbits before the SMBHs coalesce.
The main idea of \ketju{} is depicted in \autoref{fig:ketju_illustration}.

\subsection{\gadgetfour{}}

The code presented here extends the public version of \gadgetfour{}, which is described comprehensively by \citet{2021MNRAS.506.2871S}.
Here we briefly summarise the main features of \gadgetfour{} that are relevant for understanding how the \ketju{} module operates.
The modifications added to our \ketju{} version of \gadgetfour{} are discussed later in section~\ref{sec:code_interface}.

\gadgetfour{} is an $N$-body and hydrodynamics code that includes several methods for calculating gravitational interactions
and for integrating the dynamics of particles. Gas dynamics is modelled using smoothed particle hydrodynamics (SPH), and a basic
radiative cooling module, star formation and subgrid stellar feedback model \citep{2003MNRAS.339..289S} are also included.
However, the public version of \gadgetfour{} does not include models for gas accretion onto SMBHs and the associated feedback,
or models for seeding SMBHs in cosmological simulations.
Therefore in this paper we only study collisionless \ketju{} simulation applications without gas.
\gadgetfour{} supports both physical coordinate based simulations of isolated systems and periodic boxes,
as well as cosmological simulations using comoving coordinates.

The different methods available for calculating gravitational interactions are the one-sided tree-based multipole expansion,
and the fast multipole method (FMM), both of which can be paired with a particle mesh (PM) method for faster calculation of long-range interactions.
For the one-sided tree method, the particle distribution is divided using an oct-tree, and the acceleration of each particle is calculated
by walking this tree from top to bottom.
The gravitational acceleration caused by the particles within a tree node is calculated either using a multipole expansion of a set order of the
particles within the node, or by recursively considering the child nodes describing smaller volumes of space, depending on an opening criterion with a user-specified accuracy parameter.
In the FMM method the tree structure is used similarly, but instead of simply considering particle-node interactions, the method computes multipole expansions for full node-node interactions.
This results in symmetric interactions between particles, allowing manifest momentum conservation when paired with a suitable time integration scheme,
in contrast to the one-sided tree method which uses asymmetric interactions.

For all gravitational interactions, \gadgetfour{} uses a spline softening kernel with support within $2.8\epsilon$,
where $\epsilon$ is a user-specified Plummer-equivalent softening length, which can be different for different particle types \citep{2021MNRAS.506.2871S}.
Outside the support of the softening kernel, all interactions are Newtonian.

For time integration, \gadgetfour{} includes two different schemes, both based on leapfrog integration and a hierarchy of timebins.
The first scheme is the traditional nested time integration scheme, which was also used in earlier versions of GADGET \citep{2005MNRAS.364.1105S}.
In this scheme the particle timesteps $\Delta t$ are determined based on their accelerations $\vb{a}$ in the previous timestep and a user-specified accuracy parameter $\eta$ as
\begin{equation} \label{eq:gadget_timestep}
\Delta t = \sqrt{\frac{2\eta \epsilon}{|\vb{a}|}}.
\end{equation}
The timesteps are then truncated to timestep bins based on a powers-of-two division of the simulation time.
When the current simulation time is evenly divided by the stepsize of a given bin, the particles on that bin are said to be active.
Gravitational accelerations need to be calculated only for the active particles, but all particles act as sources,
requiring the full particle distribution to be included when constructing the oct-tree for gravity calculations.
The asymmetric interactions caused by this scheme also result in the non-conservation of momentum in the system.

The second scheme is the hierarchical time integration, based on splitting the Hamiltonian of the system recursively into so-called slow and fast components.
The particles are assigned timebins as above, but in this scheme only the active particles are included as sources of gravity.
This allows symmetric interactions and manifest momentum conservation when paired with FMM, and can also improve performance
for systems where the hierarchy of timebins is deep.

To parallelise calculations, \gadgetfour{} uses the Message Passing Interface (MPI) to communicate between tasks or processes
that are run in parallel on different CPU cores potentially across multiple supercomputer nodes.
Each task stores a subset of the simulation particles and is responsible for evolving them,
with the distribution of particles across tasks based on the spatial distribution of particles.

\subsection{The regularised integrator}

To integrate the dynamics of BHs and their surrounding stellar component, \ketju{} uses an updated version of
the MSTAR integrator \citep{2020MNRAS.492.4131R}.
MSTAR uses algorithmic regularisation \citep{1999CeMDA..74..287M,1999MNRAS.310..745M,1999AJ....118.2532P},
which employs a time transformation to a new time coordinate to allow integrating gravitational interactions without
running into numerical issues caused by the diverging Newtonian potential $(\Phi \propto r^{-1})$.

To reduce roundoff error when calculating gravitational interactions between nearby particles,
the code also uses a minimum spanning tree--based chained coordinate system \citep{2020MNRAS.492.4131R}.
In order to achieve tight numerical tolerances while enabling efficient parallelisation
to hundreds of CPU cores for systems of thousands of particles the code uses an MPI-parallelised Gragg--Bulirsch--Stoer extrapolation scheme.
The accuracy of the integration is controlled with a per-step relative error tolerance parameter, $\epsilon_\mathrm{GBS}$,
and gravitational forces are calculated using direct summation.

Due to the time transformation used in algorithmic regularisation,
the physical time $t$ needs to be integrated like other coordinates of the system,
and integrating the system over a specified physical time interval $\Delta t$
requires iteratively changing the stepsize  to reach the desired final time.
The accuracy of this iterative solution is controlled by the output time relative tolerance parameter $\epsilon_t$,
with the iteration ending when the desired tolerance is reached, i.e.\ when
\begin{equation} \label{eq:GBS_timetolerance}
|t-t_0|/\Delta t < \epsilon_t.
\end{equation}
Compared to the MSTAR version discussed by \citet{2020MNRAS.492.4131R},
the publicly released version presented here
also includes in addition post-Newtonian (PN) dynamics for black holes,
mergers of black holes with GW recoil kicks, optional gravitational softening between stellar particles,
and dynamic order control.

Dynamic order control allows adjusting the number of different substep divisions
used in the Gragg--Bulirsch--Stoer extrapolation algorithm, which improves the efficiency of the code particularly
in situations where integrating the system at the desired accuracy does not require a particularly
high integrator order,
e.g.\ when a system is integrated for only a small fraction of its relevant dynamical timescales.
This dynamic order control is implemented using standard methods \citep[e.g.][ch. II.9]{hairerODEBook},
with some simplifications to allow for computing multiple rows of the extrapolation table
in parallel.

The support for PN dynamics and softening divides the particles in the integrator into two types:
particles that always behave like point particles and have PN terms in their mutual interactions,
which are typically used to represent black holes,
and particles that only experience Newtonian gravity and may optionally have softened interactions
between each other, which we call stellar particles due to the typical use case.
The optional softening between the stellar particles is implemented using the same spline softening
kernel as is used in \gadgetfour{} \citep{2021MNRAS.506.2871S}, allowing for continuous transition of stellar particles
between \gadgetfour{} and the regularised integrator.
The importance of this feature will be demonstrated in Section~\ref{sec:softening_in_regions}.
The gravitational softening only affects interactions between stellar particles,
and interactions between stellar particles and black holes are always non-softened.

For PN dynamics, the expressions from \cite{1985PhRvD..31.1815T}
are used for the \pn{1} and spin terms, including all the non-linear $N$-body cross-term effects at this order,
while higher-order corrections valid for binary systems
up to \pn{3.5} order are adopted from equation~(203) of \cite{2014LRR....17....2B}.
Apart from a quadrupole spin interaction term that is specific for black holes,
the PN terms implemented in the code are also valid for other types of compact objects.
The code also supports enabling and disabling individual correction terms,
so it is possible to run simulations using  e.g.\ only the leading GW radiation reaction
term at \pn{2.5} order while ignoring the other terms, if so desired.
To integrate the equations of motion that include velocity dependent accelerations,
the code applies the auxiliary velocity scheme \citep{2010CeMDA.106..143H,2015CeMDA.121..211P},
which allows the integration scheme to remain symmetric and the extrapolation algorithm to operate correctly.

As the PN effects allow black hole binaries to shrink due to GW emission, the integrator also now includes
a merger model based on the fitting formulae from \cite{2015PhRvD..92b4022Z},
which predicts the black hole merger remnant spin, mass, and recoil velocity relative to the binary rest frame.
The binary rest frame is approximated with the \pn{3} formulae from \citet{2001CQGra..18..753D},
with the addition of the leading spin-dependent correction \citep[e.g.][]{2009PhRvD..80l4015K}.
The merger is by default performed when the binary separation is
\begin{equation} \label{eq:merger_distance}
r= 12 R_\mathrm{S} = 24 G( m_1 + m_2)/c^2,
\end{equation}
where $R_\mathrm{S}$ is the Schwarzschild radius corresponding to the total binary mass, $m_{1,2}$ are the masses of the binary components, $G$ is the gravitational constant, and $c$ is the speed of light.

The rather large merger separation is necessary as the PN formulae used in the calculations are only approximations.
At smaller separations, first the centre-of-mass formulae used for computing the rest frame begin to become inaccurate,
resulting in oscillations of the calculated centre-of-mass velocity and an incorrect final velocity of the merger remnant,
while at even smaller separations the acceleration formulae become inaccurate,
leading to clearly unphysical behaviour such as binaries reverting back to expansion.
The default merger distance of $12 R_\mathrm{S}$ has been chosen to avoid most of these issues while allowing the majority of the binary inspiral to be captured.
The distance is also user-configurable, and values as low as $6 R_\mathrm{S}$ can work fairly well at least for low-spin systems.

As the binary is merged at a wider separation than the initial condition distance of $\sim 3 R_\mathrm{S}$ used in
deriving the approximate  \cite{2015PhRvD..92b4022Z} fitting formulae, the resulting binary properties cannot be regarded as exact predictions.
In addition, the properties of the fitting functions cause the results to oscillate over the range of possible
values that can be obtained with a given spin and mass configuration during the orbit of the binary,
and thus the results can be regarded as an effectively random sample from the distribution of possible remnant properties.

The integrator code is published in its own repository as a library with a simple interface.
In addition to being used in \ketju{}, it can also be used as a stand-alone code through a simple driver program,
or called from other simulation codes.
This should allow implementing similar functionality as presented here also in other simulation codes
apart from \gadgetfour{}, such as AREPO \citep{2010MNRAS.401..791S}, RAMSES \citep{2002A&A...385..337T} and ENZO \citep{2014ApJS..211...19B}, by only implementing the interface code needed for passing particle data to and from
the integrator in a similar manner as described in the next section.

\subsection{Combining \gadgetfour{} with the regularised integrator}
\label{sec:code_interface}

\begin{figure}
\includegraphics[width=\columnwidth]{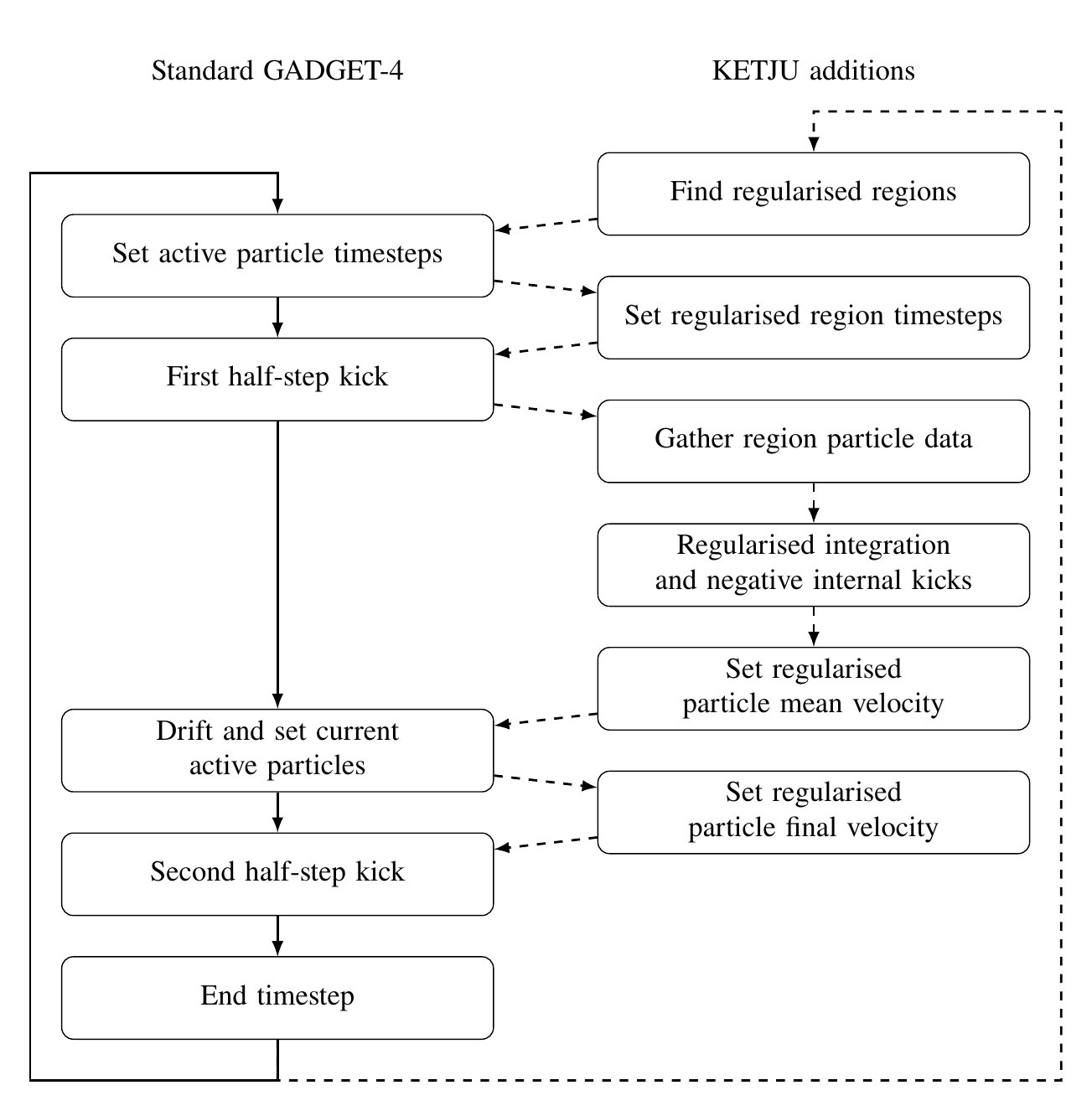}
\caption{
A schematic of the actions performed when integrating over a single leapfrog timestep.
The solid lines show the path taken in standard \gadgetfour{}, while the dashed lines
show the path taken when the \ketju{} module is active.
The figure is simplified by ignoring some tasks such as output or other additional physics
models.
}
\label{fig:ketju_flowchart}
\end{figure}

The main addition to the \gadgetfour{} code required for \ketju{} is the interface code,
which handles passing particle data to and from the integrator, as well as adjusting the
leapfrog timesteps assigned to the particles integrated
with the regularised integrator.
The main steps in the operation of the code are illustrated in \autoref{fig:ketju_flowchart}.

\subsubsection{Finding the regularised regions}

On each timestep before the first kick operation,
the code first searches for active BH particles and their surrounding stellar particles
within a spherical regularised region with a user-specified region radius, $\rreg$.
The radius of the region must be larger than the softening kernel $(\rreg \geq 2.8\epsilon)$
in order to ensure that all interactions between BHs and stellar particles are non-softened both inside and outside the regularised regions.
Overlapping regions are merged together, so that the spatial extent of each
region is a union of spheres, with a BH at the centre of each sphere as depicted in \autoref{fig:ketju_illustration}.
Only active particles are included in the regions, but the following timestep limiting
logic aims to ensure that all stellar particles are active when they enter within $\rreg$
of a BH.

Any other types of particles, such as gas and dark matter (DM), are treated as in standard \gadgetfour{}, even
if they are within $\rreg$.
The code does however include support for treating multiple user-specified types of non-gaseous particles as stellar
or BH particles when constructing the regularised regions, as there are no fundamental differences between
the different non-gaseous particle types apart from an integer tag in \gadgetfour{},
with the limitation that they must all use the same softening length.
This allows for applications where different particle types are used in additional physical models
within \gadgetfour{} to e.g.\ track newly created stellar particles,
or even including DM particles in the regularised regions so that they too
can have point-mass interactions with black holes.

The search for particles is implemented in two phases.
First each parallel task walks through the particles that are stored in its local memory
to find the BH particles.
The BH particle coordinates and other relevant properties are then communicated to all tasks,
and each task then checks for each stellar particle in its local memory if it belongs to a region
around any of the BHs.
The stellar particle data is then
communicated between tasks only just before calling the regularised integrator.
This simple approach requires only little communication between tasks during this particle search
phase, and has shown good performance in our tests, taking in general less than one percent of the
total computation time.

\subsubsection{Timestep limiting}

The timestep determination of the particles within each regularised region is altered to account for the
fact that their full dynamics are no longer integrated with the simple leapfrog integrator of \gadgetfour{}.
The timestep of each region is set by finding the lowest timestep assigned
to a particle within a configurable radius that is set by default to $100 \,\rreg$,
excluding the particles that belong to the region, and using that as the timestep of the region.
This ensures that the internal dynamics of the region are visible to the particles that are directly affected by them.
The default size of this timestep limiting radius is set quite conservatively to ensure
similar behaviour as earlier versions of \ketju{} where this limit was imposed based on all particles.
However, in general the smallest timesteps are found near the BHs just outside the regularised regions,
and the size of this timestep limiting region has little practical impact.
In addition to the particles in a regularised region, the stellar particles within a radius of $r<2\rreg$ are also
limited to the timestep of the region to ensure that they are active when they enter the region.

Additional limits on the timestep of the region are placed based on the motion of the BHs and stars within the
region, to guard against cases where the particle acceleration based timestep limit is not sufficient due to large velocities that would cause particles to pass through the regularised region without being integrated correctly.
These limits consist of:
(i) the centre of mass (CoM) acceleration limit, evaluated using equation \eqref{eq:gadget_timestep} with the softening $\epsilon$ replaced by $0.1 \rreg$;
(ii) the CoM velocity $v_\mathrm{CoM}$ based limit
\begin{equation}
\Delta t \leq \frac{0.1\, \rreg}{v_\mathrm{CoM}},
\end{equation}
which ensures that the region does not move too much relative to its size in a single timestep;
(iii) the individual BH velocity $v_{\rm BH}$ based limit
\begin{equation}
\Delta t \leq \frac{0.3\, \rreg}{v_{\rm BH}},
\end{equation}
which ensures that a BH cannot move outside the region during a single step;
and finally (iv) a limit based on the stellar velocity dispersion $\sigma$
\begin{equation}
\Delta t \leq \frac{\rreg}{6\, \sigma},
\end{equation}
which acts to ensure that essentially no stellar particle moves through the region in a single step.
The numerical factors in these limits are chosen so that particles
that enter a region do not come too close to the central BHs during the final leapfrog step before the regularised integrator takes over,
even in the case of fast relative motions.
Otherwise, the final half-step kick could be significantly affected by softening
and the effect of the BH on the motion of the particle would not be accurately captured.
The main timestep criterion based on the surrounding particle timesteps is however what typically sets the timestep,
and these limits are only intended to reduce the impact of the rare cases where it is not sufficient to capture the dynamics.

\subsubsection{Integration}

The particles in the regularised regions undergo normal leapfrog kicks,
but the contributions from the interactions within a given region
are then removed by applying additional negative kicks before and after calling the regularised integrator.
This avoids the need to modify how the kicks are performed in \gadgetfour{},
while also avoiding double counting the interactions which are already accounted for by the regularised integration.
The leapfrog kicks allow including the perturbing effects of the particles that do not belong to the regions
at the same level of accuracy as is achieved for the particles integrated by the standard \gadgetfour{} leapfrog integrator.
As the BH particles are in general the most massive particles in the applications \ketju{} is mainly intended for,
the effects of the perturbations on their motion are usually captured even more accurately than this.
This is because the region timestep is at least as small as the timesteps of the surrounding perturbing particles,
and the acceleration of the massive BH caused by the perturbing particle is smaller than
the corresponding acceleration of the much less massive perturbing particle due to the gravity of the BH.

After the first half-step kick has been applied, the particle data is passed to the regularised integrator,
and the region is integrated over the leapfrog timestep.
However, as algorithmic regularisation is used in the integration,
the timespan of the integration will not be exactly equal to the given timestep,
and will instead only match it to the tolerance specified by $\epsilon_t$.
The integrator output time tolerance parameter $\epsilon_t$ must therefore be sufficiently small,
so that the error from these slight time differences is not significant compared to the accuracy of the \gadgetfour{} leapfrog integration.
The integration can also be performed in parts to allow for output of the BH data
with a time interval that is shorter than the timestep of the region.
The BH data is then outputted in HDF5-format after all regions have been integrated.

Different regions are integrated in parallel, with the available MPI tasks allocated between the regions
based on the estimated computational cost of each region.
The computational cost estimate $C$ is based on the actual measured CPU time taken
by the region on the previous timestep,
or if not available,
on the particle count $N_\mathrm{part}$ assuming that the $\bigO(N_\mathrm{part}^2)$ gravitational interactions account
for the bulk of the cost.
Each region is then allocated $N$ tasks so that $C/N$ is approximately equal for all regions,
which should minimise the time tasks spend being idle while waiting for other regions to finish their integration.
In some cases there may not be enough tasks to integrate all regions in parallel,
in which case sets of parallel integrations are queued in sequence so that
the total estimated computational time is minimised.

For the integration, the particle data is always converted to physical coordinates in the inertial rest frame of the centre of mass
of the region, as the  integration scheme adopted in the regularised integrator does not support the comoving integration formalism.
This is not an issue for the accuracy of the dynamics even in cosmological simulations,
as compared to using comoving integration,
performing the integration in physical coordinates without additional correction terms
only ignores the effect of the cosmological constant or alternatively dark energy during the integration.
For typical systems that \ketju{} might be applied to in a cosmological context,
this ignored contribution is much smaller than other sources of error in the system,
such as the integration error or not including the PN terms in all interactions.
The inertial frame of the region is drifted over the leapfrog timestep similarly to a simulation particle,
i.e.\ either in physical or comoving coordinates as required by the type of the simulation.

In some cases, the computational cost of the regularised integration can be
significantly increased by the formation of very tight binaries,
which may not be of physical interest in all cases.
One such case is the typical configuration in galactic simulations,
where individual stellar particles have masses of around $10^5\, \Msun$,
and the rare sub-parsec scale binaries they may form with an SMBH do not meaningfully
affect the dynamics of the system, at least when compared to the computational effort
needed to integrate them.

To reduce the performance impact of such binaries, the code can optionally
add an artificial kick that expands the tightest binary in the system if it has
a short enough period to affect the integration performance,
excluding of course any SMBH binaries that are always physically meaningful.
The kick is applied before the integration, and its magnitude is chosen so that
the period of the binary is doubled.
Application of such artificial kicks naturally breaks the conservation of energy
in the system, so the choice of using this mechanism depends on the desired balance of performance and accuracy.
We enable this artificial expansion of tight binaries for our tests presented in this
paper, without any apparent impact on the physical results.
Other authors have also applied similar schemes to reduce the impact of such unimportant tight binaries,
for example \citet{2001ApJ...563...34M} directly removed such stellar particles from their simulation.

After the integration, the data in the integrator structures represents the resulting particle
state at the end of the timestep at time $t+\Delta t$,
but \gadgetfour{} has only completed
the first kick operation of this timestep.
To update the \gadgetfour{} data for the  particles in the regularised regions,
the final particle velocities are first stored in a separate field,
while the velocity field of each particle in the \gadgetfour{} particle data
is set to a mean velocity
\begin{equation}
\vb{v}_\mathrm{mean} = \frac{\vb{x}(t+\Delta t) - \vb{x}(t)}{\Delta t}
\end{equation}
that results in the particle arriving at the
correct final position ${\vb{x}(t+\Delta t)}$  at the end of the leapfrog drift step.
This avoids the need to modify the \gadgetfour{} drift routine to check which particles
have been integrated with the regularised integrator.
After \gadgetfour{} has performed the drift step,
the velocity field of the particles is updated to the correct final velocity ${\vb{v}(t+\Delta t)}$.

\subsubsection{Interaction with different \gadgetfour{} modes}

The operation of the interface code is essentially the same in both time integration
modes available in \gadgetfour{}.
It can however be better theoretically justified in the hierarchical time integration
mode, where the regularised regions behave essentially like additional fast systems
with the leapfrog evolution replaced by the more accurate regularised integration.
So instead of the evolution operator of equation~(50) of \citet{2021MNRAS.506.2871S}
\begin{equation}
\begin{aligned}
E(H,\Delta t) \simeq
&K_\mathrm{P}^\mathrm{P}\left(\frac{\Delta t}{2} \right)
K_\mathrm{F}^\mathrm{F}\left(-\frac{\Delta t}{2} \right)
\\
& \times E\left(H_\mathrm{F}, \frac{\Delta t}{2} \right)
D_\mathrm{S}(\Delta t)
E\left(H_\mathrm{F}, \frac{\Delta t}{2} \right)
\\
& \times K_\mathrm{F}^\mathrm{F}\left(-\frac{\Delta t}{2} \right)
K_\mathrm{P}^\mathrm{P}\left(\frac{\Delta t}{2} \right),
\end{aligned}
\end{equation}
where $K$ and $D$ are the kick and drift operations, and $\mathrm{P} = \mathrm{S} + \mathrm{F}$
is the set of particles split into slow (S) and fast (F) systems,
we now have
\begin{equation}
\begin{aligned}
E(H,\Delta t) \simeq
&K_\mathrm{P}^\mathrm{P}\left(\frac{\Delta t}{2} \right)
K_\mathrm{R}^\mathrm{R}\left(-\frac{\Delta t}{2} \right)
\\
&\times D_\mathrm{S}(\Delta t)
E_\mathrm{R}(\Delta t)
\\
&\times K_\mathrm{R}^\mathrm{R}\left(-\frac{\Delta t}{2} \right)
K_\mathrm{P}^\mathrm{P}\left(\frac{\Delta t}{2} \right),
\end{aligned}
\end{equation}
where $E_\mathrm{R}$ is the evolution of the regularised region particles (R)
implemented by the regularised integrator.
Note that here $D_\mathrm{S}$ and $E_\mathrm{R}$ commute, so that
$D_\mathrm{S}(\Delta t) E_\mathrm{R}(\Delta t) =
E_\mathrm{R}(\Delta t/2) D_\mathrm{S}(\Delta t) E_\mathrm{R}(\Delta t/2)$.

An important difference to the hierarchical leapfrog integration
is however the fact that the different regularised
regions do not interact with each other during the regularised integration,
whereas in the hierarchical leapfrog integration all particles at a given level
interact with each other.
However, due to the timestep limitations, the interaction between nearby but separate
regularised regions is accounted for at a sufficient level of accuracy by the perturbing
leapfrog kicks.

The different gravity calculation modes also do not affect the operation of the interface code,
as they only change how the acceleration entering the leapfrog kicks is calculated in \gadgetfour{}.
In principle the different modes might produce slightly different results for the gravitational
accelerations caused by interactions within a region, which are then removed by applying additional
negative kicks, but this is not accounted for in the interface code, which always uses direct summation
for calculating these negative kicks.
However, this is not an issue, as all the gravity calculation modes
behave like direct summation on the scales that are typical for the regularised regions.
The choice of the gravity algorithm and time integration mode can therefore
be based on their different strengths.

For the types of simulations that \ketju{} is particularly well suited for,
the momentum conservation and better handling of deep timestep hierarchies
given by the combination of hierarchical time integration and FMM gravity
seems to make it the preferred default choice.
Note however that some momentum non-conservation is introduced into the simulation
by the regularised integrator,
most importantly through the physical effects of GW emission included in the PN equations
of motion and the BH merger model,
and to a lesser degree due to numerical errors as the regularised integration scheme
is not manifestly momentum conserving.

\section{Code tests}
\label{sec:Code_tests}

\subsection{Code settings}

We run all the code tests and example applications on the Mahti supercomputer
hosted by CSC -- IT Center for Science, Finland.
The main CPU nodes of Mahti have two AMD Rome 7H12 CPUs with 64 cores each, and
256 GB of memory. For most runs we use just a single node.

Unless otherwise noted,
we use an error tolerance of $\epsilon_\mathrm{GBS}=10^{-8}$ and an output time tolerance of $\epsilon_t = 10^{-6}$
for the regularised integration.
For gravity calculations in \gadgetfour{}, we use FMM with second order multipoles,
a force error tolerance of $\alpha=0.005$ and an integration error tolerance of $\eta=0.002$.
The runs are performed using the hierarchical integration mode of \gadgetfour{}.
When analysing orbits of SMBH binaries, we use the quasi-Keplerian PN orbital elements of \citet{2004PhRvD..70j4011M},
as described in \citet{2019ApJ...887...35M}.

\subsection{Dynamical friction in a Plummer sphere}

In order to demonstrate that the new implementation of \ketju{} in \gadgetfour{} correctly captures the dynamical friction on BHs from
stellar particles, we simulate the orbital decay of an $M_\bullet = 10^8\, \Msun$ SMBH on a circular
orbit in a spherical galaxy model.
The model consists of only stellar particles distributed according to a Plummer density profile
\begin{equation}
    \rho(r) = \frac{3 M}{4\pi r_\mathrm{s}^3} \left( 1+\frac{r^2}{r_\mathrm{s}^2}\right)^{-5/2},
\end{equation}
with total mass $M=10^{11} \Msun$ and with a scale radius of $r_\mathrm{s}=3\,\mathrm{kpc}$.
The SMBH is set initially on a circular orbit at $r = r_\mathrm{vir} = 16 r_\mathrm{s} / 3\pi$.

The Plummer density profile was chosen for this test as the expected decay
rate of the orbit can be derived analytically,
and is given by \citet{2018ComAC...5....5R} as (see also \citealp{BTgaldynbook})
\begin{equation}
\dv{r}{t} = -\frac{8\pi G^2 \ln{\Lambda} \rho(r) \chi M_\bullet r}{V_c^3 \left[1+3(1 + r^2/r_\mathrm{s}^2)^{-1}\right]}.
\end{equation}
Here
\begin{equation}
V_c = \left( \frac{G M}{r}\right)^{1/2} \left( 1 + \frac{r_\mathrm{s}^2}{r^2}\right)^{-3/4}
\end{equation}
is the circular orbit velocity, and
\begin{equation}
\chi = \operatorname{erf}(X) - 2 \pi^{-1/2} X \exp(-X^2),
\end{equation}
where
\begin{equation}
X = \frac{V_c}{\sqrt{2} \sigma}
\end{equation}
and $\sigma$ is finally the velocity dispersion given by
\begin{equation}
\sigma = \sqrt{\frac{G M}{6 \sqrt{r^2 + r_\mathrm{s}^2}}}.
\end{equation}
For the Coulomb logarithm we use the value \citep[e.g.][]{BTgaldynbook}
\begin{equation}
\ln{\Lambda} = \ln{\left(\frac{b_\mathrm{max}}{b_{90}}\right)}
             = \ln{\left(\frac{2 r_\mathrm{vir}}{G m \sigma(r_\mathrm{vir})^{-2}}\right)} \approx \ln{\left(\frac{10\,\mathrm{kpc}}{35\,\mathrm{pc}}\right)} \approx 5.7,
\end{equation}
where $b_\mathrm{max}$ and $b_{90}$ are the maximal and 90 degree deflection impact parameters.

We run two \ketju{} simulations of this system at stellar particle counts of $N=10^5$ and $N=10^7$,
or stellar particle masses of $10^6 \,\Msun$ and $10^4 \,\Msun$,
resulting in SMBH to stellar particle mass ratios of $100$ and $10^4$, respectively.
These resolutions are representative of the range of resolutions that have been typically used
in simulations of SMBH dynamics within galaxies.
For the $N=10^7$ particle run we use a softening length of $\epsilon=50\,\mathrm{pc}$
for interactions between stellar particles and a
regularised region radius of $\rreg=3\epsilon=150\,\mathrm{pc}$, while for the $N=10^5$ particle run
we use a softening length of $\epsilon=100\,\mathrm{pc}$ and a regularised region radius of
$\rreg=3\epsilon=300\,\mathrm{pc}$.

We also run the same runs with standard \gadgetfour{} to allow for a comparison of
its behaviour to \ketju{}.
For these comparison runs we also include a version of the $N=10^7$ particle system
using a small softening length of $\epsilon=5\,\mathrm{pc}$,
to show how using \ketju{} compares to simply increasing the resolution in standard \gadgetfour{}.
In the \gadgetfour{} runs the interactions between stellar particles and the SMBH
are softened with the same softening length as the interactions between stellar particles,
while in the \ketju{} runs the SMBH always uses non-softened gravity.

\begin{figure}
\includegraphics{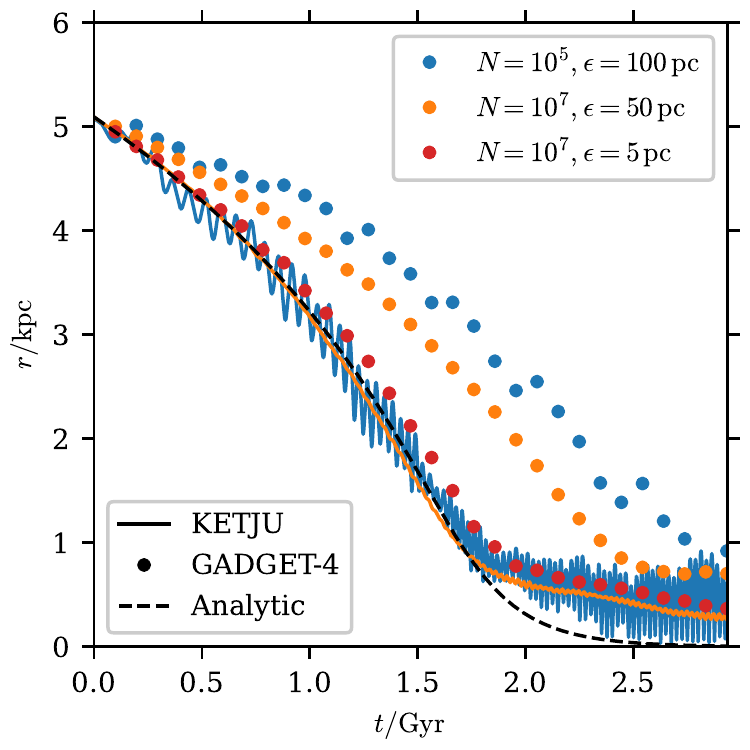}
\caption{
Sinking of an SMBH in a Plummer sphere of stellar particles due to dynamical friction.
Simulations are run using different stellar particle counts $N$ and gravitational softening lengths
$\epsilon$, using both \ketju{} and standard \gadgetfour{}.
The results are compared to an analytic prediction, which agrees well with the
\ketju{} results as well as the \gadgetfour{} run with $\epsilon=5\,\mathrm{pc}$.
}
\label{fig:BH_plummer_sinking}
\end{figure}

The results are shown in \autoref{fig:BH_plummer_sinking}.
Both \ketju{} runs follow the analytic prediction very well until around $r=1\,\mathrm{kpc}$,
apart from some oscillations due to deviations from a circular orbit particularly at lower
particle mass resolutions.
The deviation from the analytic prediction at small radii is expected due to the
assumptions of the analytic model breaking down, and is seen also in tests of
other codes using this setup \citep[e.g.][]{2018ComAC...5....5R,2021ApJ...916....9M}.
Apart from the oscillations which are made more significant due to the increased stochasticity
at lower particle counts, the dynamical friction captured by \ketju{} does not appear to
depend on the resolution at these particle counts.
There is also no apparent dependency on the softening length,
as expected since the interactions with the SMBH are non-softened.

The standard \gadgetfour{} results on the other hand clearly deviate from the
analytic prediction, with the deviation being stronger for larger softening lengths.
This is expected, as the softened gravity for the SMBH
prevents capturing close interactions with stellar particles and
thus reduces the strength of the frictional force.
The $\epsilon=5\,\mathrm{pc}$ softening length run produces results that are in reasonably good agreement
with the analytic and \ketju{} results,
but the simulation took about twice as long to complete as the $\epsilon=50\,\mathrm{pc}$ softening length \ketju{} run.

\begin{figure}
\includegraphics{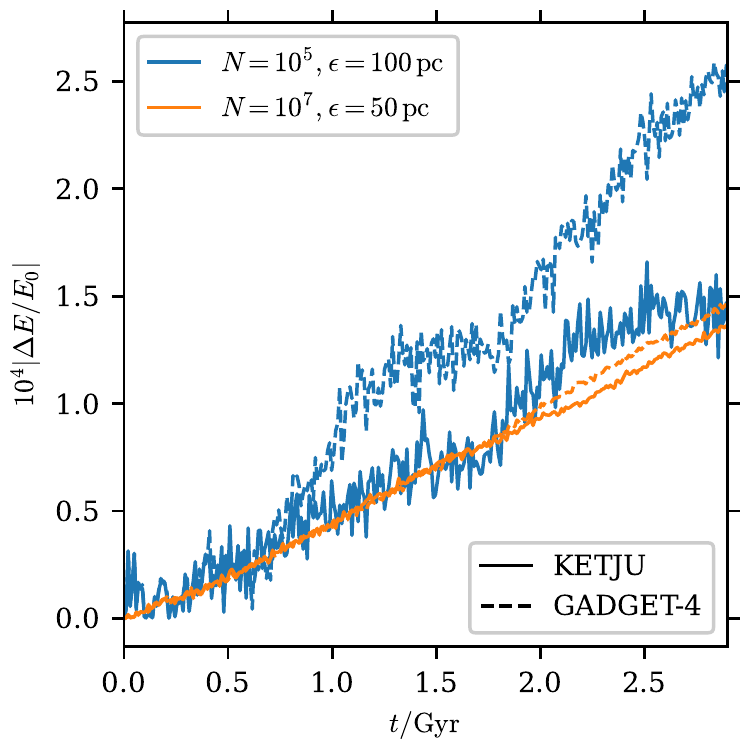}
\caption{
Relative error in the total energy of the same simulated systems as in \autoref{fig:BH_plummer_sinking}.
The total energy is conserved on the level of $\sim 10^{-4}$ for both \ketju{} and standard \gadgetfour{}, with
the energy conservation being marginally better in the \ketju{} runs.
}
\label{fig:BH_plummer_sinking_energy}
\end{figure}

We also use this test setup to study the conservation of energy in the system.
The relative error in the total energy of the system for the runs is shown in \autoref{fig:BH_plummer_sinking_energy}.
The total energy of the system decreases slightly with a final relative error of about $10^{-4}$
in both the standard \gadgetfour{} as well as
the \ketju{} runs, but the energy conservation is still somewhat better when the system is run with \ketju{}.
The relatively small effect of \ketju{} on the total energy conservation is expected,
as only a small fraction of the stars have strong interactions with the SMBH where
the more accurate integration in \ketju{} improves the energy conservation.
The momentum conservation of the simulation is not measurably affected by \ketju{} in this test.

The test presented here used only stellar particles, but the results apply equally well
to interaction with DM particles at the same mass resolutions, as the regularised integrator
can be set to treat them in the same way as stellar particles.
However, including DM in the regularised regions generally requires that it uses a similar
mass resolution and the same softening length as the stellar particles.
This is because too massive particles would compromise the SMBH dynamics,
and due to the issue discussed in Section~\ref{sec:softening_in_regions} interactions between the non-SMBH
particles usually need to be softened, with the current version of the regularised integrator only
supporting a single softening length.

Using such a high resolution for DM is often not feasible due to the large number
of required DM particles.
In those cases \ketju{} would only resolve the dynamical friction from stellar particles,
while the interaction with DM would be integrated with the standard \gadgetfour{} integrator,
causing it to be underestimated.
For typical simulations this is likely not a significant problem, as the DM fractions in the centres
of galaxies where the SMBHs are typically found are rather low \citep[e.g.][]{2013MNRAS.432.1709C}, and most of the dynamical
friction comes from the stellar particles.
However, it might also be possible to pair \ketju{} with some alternative scheme for
treating the unresolved dynamical friction from DM \citep[e.g.][]{2015MNRAS.451.1868T,2019MNRAS.486..101P,2023MNRAS.519.5543M}
to improve the accuracy in these cases.
In hydrodynamical simulations,
the approach used in \ketju{} naturally cannot help in resolving the dynamical friction
from gas \citep{1999ApJ...513..252O}, and alternative approaches would need to be developed in addition.

\subsection{SMBH binary hardening rate}
\label{sec:binary_hardening}

\begin{table*}
    \centering
    \caption{Resolution parameters of the galaxy models used in determining the binary hardening convergence.
        The subscripts $\star$ and $\mathrm{DM}$ denote stellar and DM particle properties, respectively,
        with $m$ being the individual particle mass, $N$ the total number of particles,
        $\epsilon$ the softening length.
        $M_\bullet$ is the mass of the central SMBH, and the final column gives the resolution of the simulation
        as defined by the logarithmic mass ratio between the SMBH and stellar particles.}
    \label{tab:hernquist_pars}
    \begin{tabular}{ccccccc}
        \hline
        $m_\star / (10^5 \Msun)$ & $\log_{10}\left(N_\star\right)$ & $\epsilon_\star/\mathrm{pc}$ & $m_\mathrm{DM} / (10^7 \Msun)$ & $\log_{10}\left(N_\mathrm{DM}\right)$ & $\epsilon_\mathrm{DM}/\mathrm{pc}$ & $\log_{10}\left(M_\bullet/m_\star\right)$\\
        \hline
        0.5 & 6.70 & 5.0 & 1.0 & 6.66 & 300.0 & 4.8 \\
        1.0 & 6.40 & 5.0 & 2.0 & 6.36 & 300.0 & 4.5 \\
        2.0 & 6.10 & 5.0 & 4.0 & 6.05 & 300.0 & 4.2 \\
        5.0 & 5.70 & 5.0 & 10.0 & 5.65 & 300.0 & 3.8 \\
        10.0 & 5.40 & 10.0 & 20.0 & 5.36 & 600.0 & 3.5 \\
        20.0 & 5.10 & 10.0 & 40.0 & 5.04 & 600.0 & 3.2 \\
        50.0 & 4.70 & 10.0 & 100.0 & 4.60 & 600.0 & 2.8 \\
        100.0 & 4.40 & 20.0 & 200.0 & 4.30 & 1000.0 & 2.5 \\
        \hline
    \end{tabular}
\end{table*}

While the particle mass resolution does not significantly affect how well dynamical friction is resolved,
it is well known from previous studies that this can have an effect on the hardening rate of SMBH binaries,
with the hardening rate converging only at high enough mass resolutions \citep[e.g.][]{2006ApJ...642L..21B,2011ApJ...732...89K,2017MNRAS.464.2301G}.
To investigate the convergence of the SMBH binary hardening rate in \ketju{} simulations,
we run a series of idealised, isolated galaxy mergers without gas.

Each galaxy is modelled as a multicomponent sphere consisting of a $M_{\star}=2.5\times10^{11}\Msun$ stellar bulge,
a $M_\mathrm{DM}=4.9\times10^{13}\Msun$ DM halo, as determined by the $z=0$ scaling relation of \citet{2019MNRAS.488.3143B},
and a single SMBH of $M_{\bullet}=3.0\times10^9 \,\Msun$, which is consistent with observations \citep[e.g.][]{2019ApJ...876..155S}.
Both the stellar and DM mass distributions follow a Hernquist profile
\begin{equation}
\rho(r) = \frac{M r_\mathrm{s}}{2 \pi r (r + r_\mathrm{s})^3}
\end{equation}
with scale radii $r_\mathrm{s,\star}=3.0\,\mathrm{kpc}$ and $r_\mathrm{s,DM}=300\,\mathrm{kpc}$, respectively.

To create the merging system, a pair of these galaxies is set on
a near-radial (eccentricity $e\sim0.99$) Keplerian orbit with an initial separation of $r_\mathrm{sep}\sim 20\,\mathrm{kpc}$ and first pericentre passage distance of $r_\mathrm{peri}\sim8\,\mathrm{kpc}$.
We run mergers using eight different mass resolutions $M_\bullet/m_\star$, provided in \autoref{tab:hernquist_pars}.
We create four realisations of the galaxy model per mass resolution, using different pseudo-random number generator seeds when sampling the particle phase space distribution.
We then merge each combination of the model galaxy realisations at a given mass resolution, resulting in ten unique mergers per mass resolution.
The runs with different mass resolutions use different stellar and DM particle softening lengths,
chosen so that the number of stellar particles within the regularised region radius, $\rreg = 3 \epsilon_\star$, does not grow larger than a few thousand in order
to keep the computational cost at a reasonable level.
Each merger simulation is run for $500\,\mathrm{Myr}$, which is long enough for the SMBH binary to shrink to a parsec-scale separation.

To quantify the binary hardening rate, we compute the inverse semimajor axis $1/a$ and the corresponding hardening constant $H$ \citep{1996NewA....1...35Q}:
\begin{equation}\label{eq:quinlan_hardening}
    \frac{\mathrm{d}}{\dd{t}}\left(\frac{1}{a}\right) = H \frac{G\rho}{\sigma},
\end{equation}
where $a$ is the binary semimajor axis, $\rho$ is the stellar density and $\sigma$
is the stellar velocity dispersion within the influence radius $r_\mathrm{h}$ of the binary.
The influence radius is defined so that the enclosed stellar mass $M_\star(r_\mathrm{h})$ satisfies $M_\star(r_\mathrm{h}) = 2 M_\mathrm{binary}$.
To determine the value of $H$, we fit a linear function to $1/a$  in the range
between the hardening radius $a=a_\mathrm{h}=r_\mathrm{h}/16$ \citep[e.g.][]{2013degn.book.....M}
and $a=10\,\mathrm{pc}$,
which occurs before GW emission becomes the primary hardening mechanism of the binary.
For each simulation, we use the median value of $\rho/\sigma$ determined from the set of snapshots after $a=a_\mathrm{h}$.
This ensures that our method is not sensitive to fluctuations in $\sigma$ at early times when the merger remnant has not yet reached
an equilibrium.

\begin{figure}
    \includegraphics{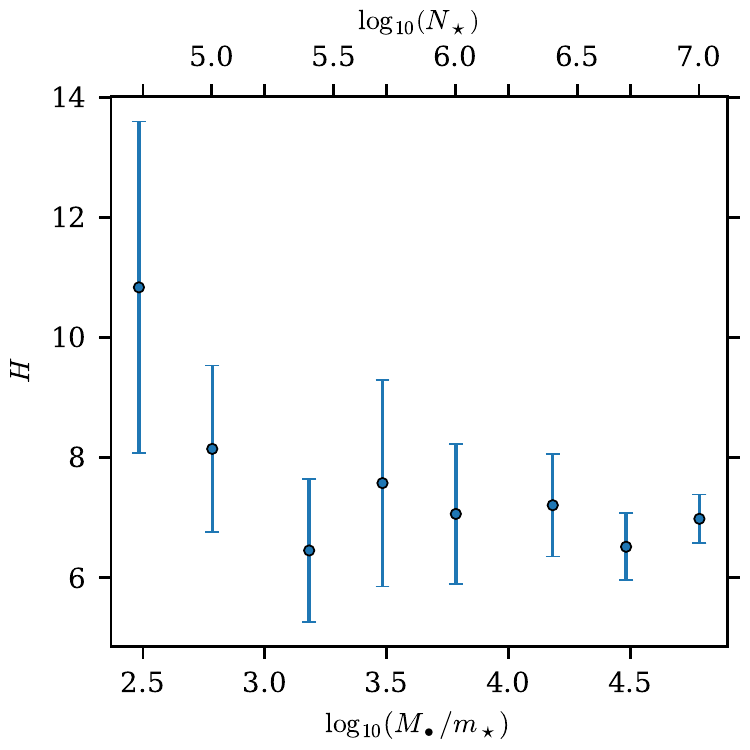}
    \caption{
    The mean and standard deviation of the hardening parameter $H$ as a function of mass resolution. The spread in the data decreases with increasing mass resolution.
    }
    \label{fig:hardening_convergence}
\end{figure}

The convergence of the hardening parameter $H$ with increasing mass resolution can be easily seen from \autoref{fig:hardening_convergence},
with the mean value of the set of runs at each resolution being nearly constant at high mass resolutions.
To more rigorously quantify the convergence of the hardening rate,
we use a statistical test to see when the distributions of $H$ from the sets of runs at different resolutions have become indistinguishable from each other.
We thus wish to investigate the following test hypothesis:
\begin{equation*}
    \begin{cases}
        H_\mathrm{null}:& \text{the variance between two sets is indistinguishable
        } \\
        H_\mathrm{alt}:& \text{the variance between two sets is not indistinguishable}
    \end{cases}
\end{equation*}

To do this, we perform a permutation test for each pair combination $(A,B)$ of mass resolutions.
The test is detailed in \citet{EfroTibs93}, with the main steps briefly outlined here.
We collect the observations from set $A$ and the observations from set $B$ into a single vector $V$ of length $N_T$, where $N_T$ is equal to the sum of the lengths of $A$ and $B$, $N_T = N_A+N_B$.
We determine the variance of the first $N_A$ elements in $V$, labelling this as $\sigma_A^2$, and the variance of the final $N_B$ elements in $V$, labelling this as $\sigma_B^2$.
As a test statistic, we calculate the logarithm of the ratio of these two variances, namely:
\begin{equation}
    t_\mathrm{stat} = \log_{10}\left(\frac{\sigma_A^2}{\sigma_B^2}\right).
\end{equation}
The test statistic is $t_\mathrm{stat}=0$ if the variances of the two samples are indistinguishable, i.e.\ when $\sigma_A^2 = \sigma_B^2$.
We build a distribution of this test statistic by shuffling the elements in $V$ without replacement $N_\mathrm{reps}=10^4$ times, generating a series of test statistics $t_{\mathrm{stat},i}$.
For each permutation of $V$, we determine the variance in the first $N_A$ elements, and the variance in the final $N_B$ elements, and calculate $t_\mathrm{stat}$.
This allows us to assess how likely the original value of $t_\mathrm{stat}$ is against random realisations of the values in sets $A$ and $B$.
To calculate the probability of observing the original value of $t_\mathrm{stat}$ under the null hypothesis, we use the two-tailed $p$-value:
\begin{equation}
    p = \frac{2}{N_\mathrm{reps}} \min\left(\sum_i \left[t_{\mathrm{stat},i} < t_\mathrm{stat}\right], \sum_i \left[t_{\mathrm{stat},i} > t_\mathrm{stat}\right]\right),
\end{equation}
where $[q] = 1$ if $q$ is true and 0 otherwise,
and the $\min$-function is used as we do not assume symmetry of the distribution
of $t_{\mathrm{stat},i}$ about some location parameter (e.g. the mean in the case of a normal distribution).
We must then decide to what level of confidence $\alpha$ we will either reject or accept the null hypothesis.
We follow established convention and choose $\alpha=0.001$, i.e. the 99.9\% confidence interval.
Following the standard procedure of hypothesis testing, if $p<\alpha$, then we reject the null hypothesis in favour of the alternate hypothesis: the variance in hardening rate in set $A$ is not indistinguishable from the variance in the hardening rate in set $B$ to the confidence level given by $\alpha$.
Conversely, if $p>\alpha$, then we cannot reject the null hypothesis, and thus must instead accept the null hypothesis.

\begin{figure}
    \includegraphics{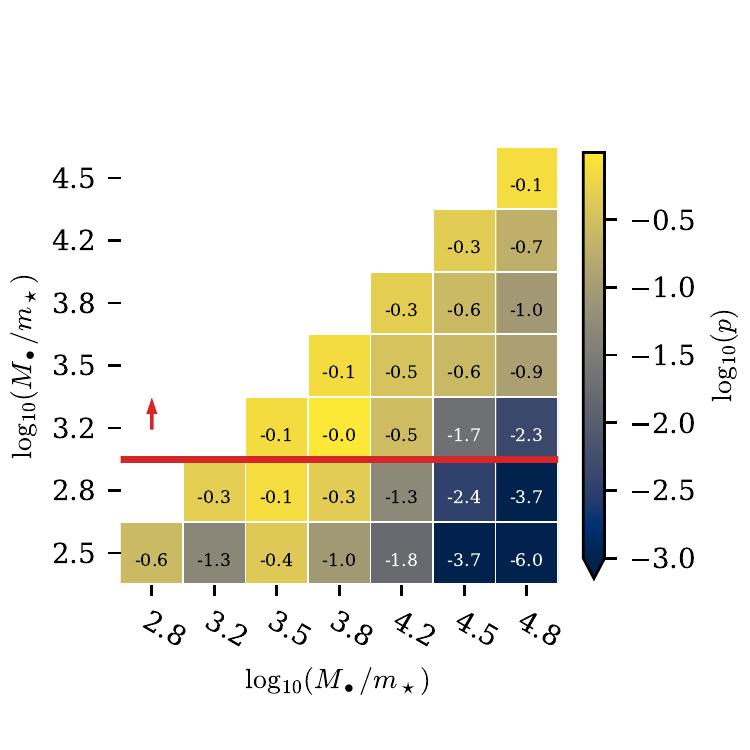}
    \caption{
    Two sample permutation test of the ratio of variances between samples. Points are coloured by the permutation test $p$-value.
    A mass resolution $\log_{10}\left(M_\bullet/m_\star\right) \gtrsim 3.0$, marked with the red horizontal line, is required for consistent convergence.
    }
    \label{fig:hardening_pvals}
\end{figure}

In \autoref{fig:hardening_convergence} we show the mean and standard deviation of $H$ for each mass resolution,
while \autoref{fig:hardening_pvals} shows the $p$-value from the permutation test.
The $p$-value for each combination of sets $A$ and $B$ is shown in the respective cell in \autoref{fig:hardening_pvals}, and
as indicated in the figure a mass resolution of $\log_{10}\left(M_\bullet/m_\star\right) \gtrsim 3$ is required for the variance between mass resolution samples to become consistently indistinguishable, and thus satisfy our convergence criterion.
By consistently indistinguishable, we mean that for a set $A$, every set $B$ of higher mass resolution has a $p$-value less than the significance level $\alpha$.
Some lower resolution samples (e.g.\ $\log_{10}\left(M_\bullet/m_\star\right) \sim 2.5$) also satisfy the convergence criterion when compared to other low resolution samples (e.g.\ $\log_{10}\left(M_\bullet/m_\star\right) \sim 3.5$), but fail to satisfy the criterion when compared to higher resolution samples (e.g. $\log_{10}\left(M_\bullet/m_\star\right) \sim 4.5$): thus at this mass resolution, consistent convergence is not achieved.

\begin{figure}
    \includegraphics{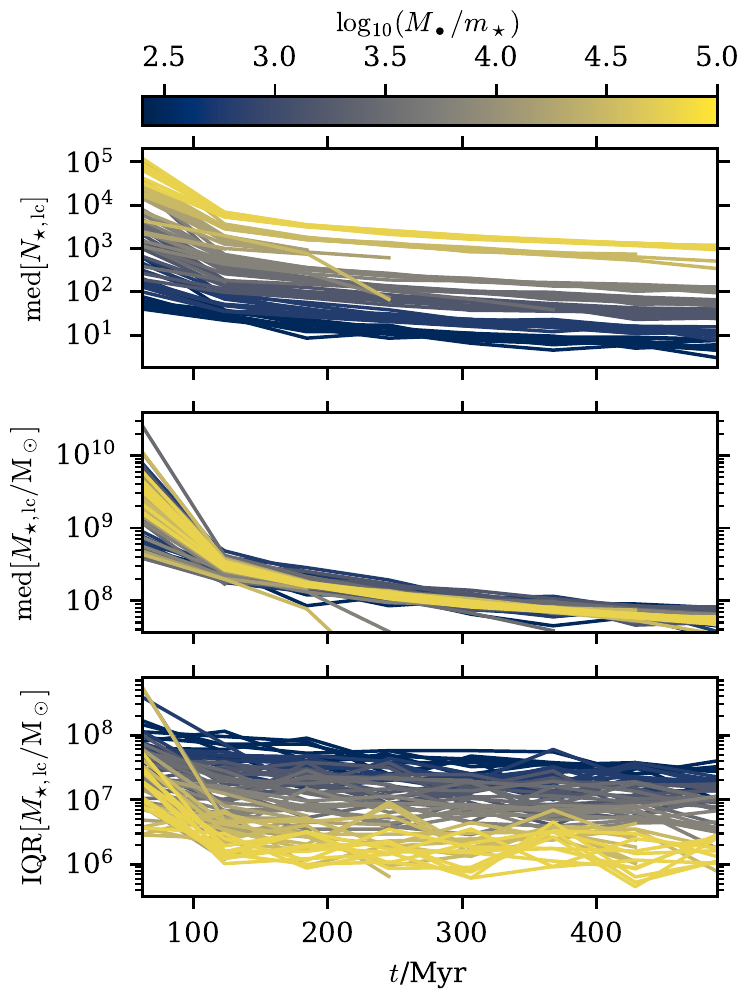}
    \caption{
        \textit{Top}: Number of stellar particles within the loss cone as a function of time since the SMBH binary forms a bound system. The number of particles inside the loss cone dramatically decreases with decreasing mass resolution. As a result of poor mass resolution, there are prolonged periods of an almost-emptied loss cone, effectively stalling the SMBH binary hardening.
        \textit{Middle}: Median stellar mass within the loss cone for the same time period as the top panel for each simulation, sampled over ten possible binary phase-space locations.
        \textit{Bottom}: The interquartile range (IQR) of stellar mass within the loss cone for the same time period as the top panel for each simulation. The IQR is determined using the same sampling method as for the median in the middle panel. Whilst the median stellar mass within the loss cone is consistent between mass resolutions, the scatter shows a gradient of higher mass resolutions corresponding to a reduced scatter. In other words, the mass entering and exiting the loss cone varies more smoothly with increased mass resolution, as expected.
    }
    \label{fig:loss_cone}
\end{figure}

In order to better understand the hardening convergence for different mass resolutions, it is informative to investigate the number of stellar particles that are able to interact with the SMBH binary: namely, those stellar particles that reside within the loss cone of the SMBH binary.
A particle in the loss cone has a specific angular momentum $J$ that satisfies:
\begin{equation}
    J < J_\mathrm{lc} = \sqrt{2 G M_\mathrm{bin} \mathcal{K} a },
\end{equation}
where $\mathcal{K}$ is a dimensionless constant of order unity.
We set $\mathcal{K}=1$ following \citet{2017MNRAS.464.2301G}.

Although the SMBH binary is expected to remain nearly stationary at the centre of the galaxy in the real Universe,
at mass resolutions that can be used in simulations the individual interactions with stellar particles
are strong enough to cause the binary to undergo Brownian motion around the centre \citep[e.g.][]{2016MNRAS.461.1023B}.
This motion affects the number of stellar particles within the binary loss cone.
To study how the random motion of the binary affects the number of particles in the loss cone,
we calculate for each simulation snapshot the number of particles in the loss cone with respect to the actual position and velocity of the binary, and in addition
with respect to ten additional possible positions and velocities where the binary could equally well have been located due to the Brownian motion.
These additional positions and velocities are uniformly sampled on spheres around the stellar CoM,
located using the shrinking sphere method \citep{2003MNRAS.338...14P},
with radii corresponding to the actual distance and relative velocity of the binary to the stellar CoM.

The median number of stellar particles within the binary loss cone across the sampled possible binary positions,
and the corresponding median stellar mass are shown in the top two panels of \autoref{fig:loss_cone} for each of the tested mass
resolutions as a function of time since the SMBHs formed a bound binary system.
We use the interquartile range (IQR) as a measure of the spread of the number of stars in the binary loss cone at each time bin, as
shown in the bottom panel of \autoref{fig:loss_cone}.
We find that the stellar mass within the loss cone is consistent between mass resolutions,
with higher mass resolution simulations ($\log_{10}\left(M_\bullet/m_\star\right) \gtrsim 4.0$) displaying less scatter as compared to lower mass resolution simulations ($\log_{10}\left(M_\bullet/m_\star\right) \lesssim 3.0$),
as well as displaying less overall sensitivity to the Brownian motion of the binary.
This resolution dependence is visible as the vertical colour gradient in the bottom panel of \autoref{fig:loss_cone}.

Critically, we find that there are prolonged periods where very few stellar particles reside within the SMBH binary loss cone for low mass resolution simulations, whereas the loss cone is never depleted for high mass resolution systems.
An emptied loss cone corresponds to an absence of stellar particles that are able to undergo hard scattering interactions with the SMBH binary, thus stalling the hardening of the binary.
Indeed, having only a few stellar particles within the loss cone dramatically decreases the efficiency with which the binary can impart orbital energy and angular momentum to the stellar particles, introducing large stochasticity in the rate at which the binary hardens.
Ultimately, this stochasticity in hardening rate manifests itself as the scatter observed in \autoref{fig:hardening_convergence} for low mass resolutions.
With an increased particle count, the higher mass resolution simulations have a steadier flow of particles that are able to harden the SMBH binary than their low resolution counterparts, resulting in reduced scatter in the hardening rate $H$.

\subsection{Loss cone refilling in a spherical system}

\begin{figure}
    \includegraphics[width=0.4\textwidth]{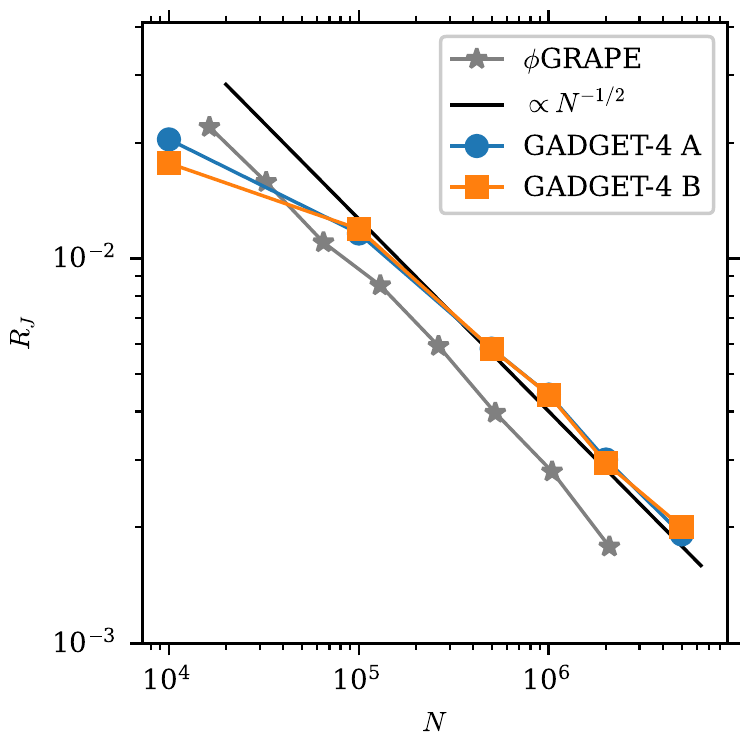}
    \caption{
    The loss cone refilling parameter $R_J$ in an isolated Hernquist sphere as a function of different particle counts $N$ using different \gadgetfour{} configurations (see the text for details).
    In addition to the \gadgetfour{} results,
    the figure also shows the results for the same test system using the $\phi$GRAPE-code,
    extracted from figure~A1 of \citet{2017MNRAS.464.2301G},
    as well as the expected $R_J \propto N^{-1/2}$ scaling.
    }
    \label{fig:lc_refilling}
\end{figure}

To supplement the hardening rate convergence test in realistic systems presented in the previous section,
we perform here in addition an idealised test to check how accurately the large-scale gravitational dynamics of \gadgetfour{} capture the processes that refill the loss cone.
We repeat the test of \citet{2017MNRAS.464.2301G} for loss cone refilling in an isolated Hernquist sphere, where loss-cone refilling mainly occurs due to two-body relaxation.
This makes the refilling rate sensitive to the accuracy of the gravitational interactions far from the centre of the galaxy.
Critically, \citeauthor{2017MNRAS.464.2301G} noted that in this test GADGET-2 did not produce converged results with the expected $R_J \propto N^{-1/2}$ scaling observed with other codes,
where $N$ is the number of stellar particles.
The authors attributed this to individual random large force errors caused by the gravity algorithms adopted in GADGET-2, and here we wish to confirm whether such errors are also present in \gadgetfour{}.

The test setup consists of a single Hernquist stellar sphere,
which we set to have a mass of $M=10^{10} \,\Msun$ and a scale radius of $r_\mathrm{s} = 1\,\mathrm{kpc}$.
We however note that the system can always be scaled to a dimensionless unit system with $M=r_\mathrm{s}=G=1$,
so that the specific parameter values are not important here.
Following \citet{2017MNRAS.464.2301G}, the softening length is set to
\begin{equation}
\epsilon = \left(\frac{2\pi}{N}\right)^{1/3}\,\mathrm{kpc},
\end{equation}
although in \gadgetfour{} this results in slightly less softened interactions compared to the \citeauthor{2017MNRAS.464.2301G} simulations which use Plummer-softening.
The system is then evolved for approximately $1\,\mathrm{Gyr}$ using different combinations of code options.
We store 50 snapshots during the evolution, and evaluate from these snapshots the refilling parameter
\begin{equation}
    R_J = \frac{N_J}{N},
\end{equation}
where $N_J$ is the total number of unique stellar particles that enter the loss cone during the simulation,
i.e.\ have their angular momentum reduced to $J<J_\mathrm{lc}$.
To evaluate the angular momentum with respect to the galactic centre,
we find the location of the centre using the shrinking spheres method for each snapshot.
For the loss cone angular momentum we use the value of $J_\mathrm{lc}=0.005 \sqrt{G M r_\mathrm{s}}$, thus matching figure~A1 of \citet{2017MNRAS.464.2301G}.

In \autoref{fig:lc_refilling} we show the results of this test with different \gadgetfour{} code configurations, compared against the results of \citet{2017MNRAS.464.2301G}.
Configuration A is the same as is used for most other runs presented in this paper, i.e. using FMM with second order multipoles, hierarchical time integration, and force and integration error tolerances set to $\alpha = 0.005$ and $\eta=0.002$, respectively.
Configuration B uses the same tolerances, but uses instead the one-sided tree calculation of gravity with first order multipoles together with the traditional nested time integration scheme, mimicking thus the configuration used
by the \gadgetthree{} version of \ketju{} that has been used in previous \ketju{} studies.

Both configurations display nearly identical results, following the expected
$R_J \propto N^{-1/2}$ scaling observed by \citet{2017MNRAS.464.2301G}, except at the very lowest resolutions.
The $R_J$-values are however slightly larger than those obtained by \citeauthor{2017MNRAS.464.2301G} using the $\phi$GRAPE-code for the same setup,
which might be due to the different treatment of the gravitational softening.
In any case, we see no evidence of the type of clearly erroneous behaviour that \citeauthor{2017MNRAS.464.2301G} observed to occur when using GADGET-2.
The loss cone refilling rate appears to be correctly captured using our reference code configuration, and in addition the rate is not very sensitive to the specific code options.

We have also performed this analysis using different force and integration error tolerances, and found only slight differences when using reasonable tolerance parameters between the default values recommended for general \gadgetfour{} simulations and tolerances up to ten times smaller than the reference tolerances adopted here.
In addition, performing this analysis for the merger runs of Section~\ref{sec:binary_hardening} showed that the refilling parameter is essentially independent of the system mass resolution, with this result again being consistent with the results of \citet{2017MNRAS.464.2301G}.

\subsection{Stellar softening in the regularised integrator}
\label{sec:softening_in_regions}

While the regularised integrator can integrate systems without requiring softening,
when combined with a large-scale \gadgetfour{} simulation it is often necessary to
include the same softening as used by \gadgetfour{}
in the interactions between stellar particles also in the regularised integration.
Otherwise a stellar particle moving in or out of a regularised region would experience
a sudden jump in potential energy as its interaction with the other stellar particles
in the region changes between softened and non-softened.
The energy error from such jumps can be
particularly significant when the region contains a relatively large mass of stellar particles
compared to the central SMBH.
Such energy errors can lead to the formation of artificial stellar cusps, while
including stellar softening in the regularised integration avoids such spurious effects.

\begin{figure}
\includegraphics{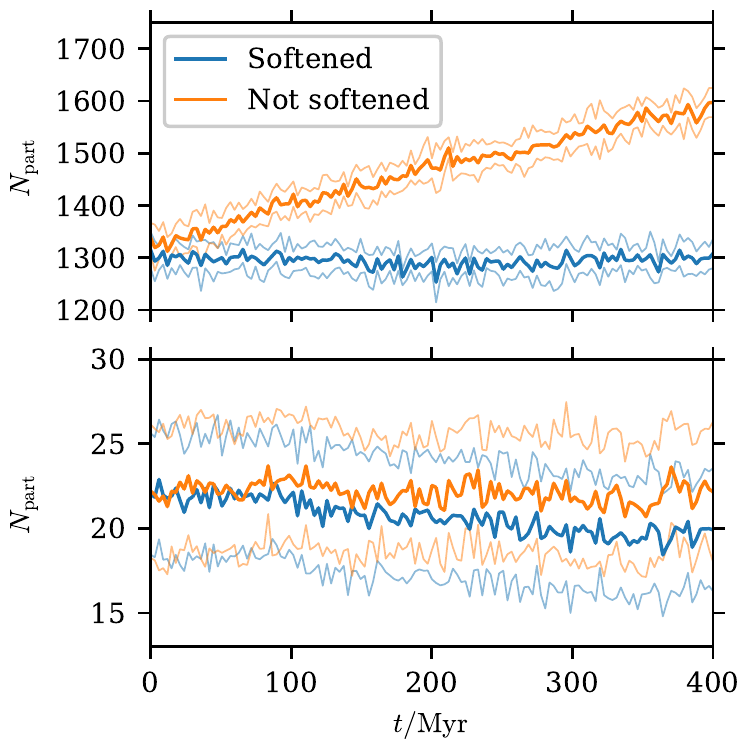}
\caption{
The evolution of the number of particles within the regularised region in simulations run
with (blue lines) and without (orange lines) softening between stellar particles in the regularised integrator.
Simulations are run with either $\rreg = 90\,\mathrm{pc}$ and $M_\bullet = 3\times 10^8\, \Msun$ (top panel) or $\rreg = 9\,\mathrm{pc}$ and $M_\bullet = 3\times 10^9 \,\Msun$ (bottom panel).
The data is smoothed over a small time window of $3\,\mathrm{Myr}$ to reduce the stochastic oscillation of the particle count,
with the thick lines showing the mean and the thin lines showing the $\pm 1\sigma$ limits.
When softening is not used, the density rapidly increases in the large region run in the top panel due to energy errors when
particles enter and exit the region, leading to the formation of an artificial stellar cusp.
In the small region run in the bottom panel this effect is not clearly apparent, as the SMBH dominates the dynamics within the region,
and the energy errors in stellar interactions are thus insignificant.
}
\label{fig:softening_in_integrator}
\end{figure}

This effect is demonstrated in \autoref{fig:softening_in_integrator}, which
shows the number of particles within the regularised region during a $400\,\mathrm{Myr}$
simulation of a Hernquist sphere galaxy model with a stellar mass of
$M_\star = 3\times 10^{11} \,\Msun$ resolved using two million particles, hosting
a $M_\bullet = 3 \times 10^8 \,\Msun$ SMBH at its centre.
The stellar softening is set to $\epsilon=30\,\mathrm{pc}$ for this run, with a regularised
region radius of $\rreg=90\,\mathrm{pc}$.
Due to this large region radius, the region initially contains about 1300 stellar particles, corresponding to a total mass of about $2\times10^8 \,\Msun$, which is comparable to the mass of the SMBH.
The simulation is run both with and without stellar softening in the regularised
integration.
The run without softening shows a rapid increase in the central density due to
the energy errors explained above, while the density in the run with softening remains approximately stable.
Similar effects would also appear if the regularised region was allowed to be
made smaller than the softening kernel,
due to the interactions between the SMBH and stellar particles then also
discontinuously transitioning between softened and non-softened forms.
To avoid these issues, all the other runs in this paper use softening between
stellar particles in the regularised regions.

For comparison, we also run simulations with a $\epsilon=3\,\mathrm{pc}$ softening,
a $\rreg=9\,\mathrm{pc}$ regularised region radius, and an SMBH with $M_\bullet = 3\times 10^9 \,\Msun$.
These settings result in the region containing about $3\times 10^6 \,\Msun$ mass of stellar particles.
In this case the effect of the energy errors appears insignificant,
which is due to the SMBH completely dominating the gravitational dynamics within
the regularised region.
The difference in the region particle count between the runs is only $\sim 2 \textnormal{--} 3$ particles,
which is comparable to the random variation during the simulations,
and could be caused purely by random variations between the runs.
These runs resemble the configuration used in some earlier simulations run with \ketju{} \citep[e.g.][]{2017ApJ...840...53R,2018ApJ...864..113R},
and thus demonstrate that in such situations the error from not using softening within the regularised integrator is unimportant.

The approach of using the same softening for interactions between stellar particles in the
regularised integrator as is used in the main \gadgetfour{} simulation is the simplest
solution to avoiding energy errors when particles enter and exit the regularised
region, but it has some potential issues for some applications.
For instance, in very high-resolution simulations where the stellar particles have physically realistic
masses, it might be desirable to integrate the regularised regions without any softening
to capture relaxation processes in the stellar clusters around SMBHs.
Depending on the softening length used on the \gadgetfour{} side and the other details of the system,
the energy error in the transition to non-softened interactions might still be large enough to compromise the results of such a simulation.
Thus, some other kind of scheme for smoothly transitioning between softened and non-softened interactions might be needed for such applications.
The current implementation also prevents including multiple particle types with different softening
lengths in the regularised regions, which might be desirable in some situations.
However, for most applications of \ketju{} that we can foresee,
the current approach of handling
gravitational softening in the regularised regions should be satisfactory.

\subsection{Integrator tolerance and PN corrections}

When running simulations with the PN-corrections enabled, the integrator tolerance
$\epsilon_\mathrm{GBS}$ plays an important role in capturing the effects
of the corrections on the binary orbit.
To test the effects of the integrator tolerance on the evolution of a binary using
the full \pn{3.5} equations of motion implemented in the code, we integrate an SMBH binary
in isolation until merger using a range of integration tolerances.

The binary components have masses of $M_1 = 10^9\ \Msun$ and $M_2 = 5\times 10^8 \ \Msun$,
and the initial orbit has a semimajor axis of $a = 200 G (M_1 + M_2)/c^2 \approx 0.03\,\mathrm{pc}$
and an eccentricity of $e = 0.1$. Starting from this initial condition it takes about
$5.3\times 10^5\,\mathrm{yr}$ for the binary to merge due to GW emission,
during which time it completes tens of thousands of orbits.
The output time relative tolerance was fixed to $\epsilon_t = 10^{-10}$ for these runs
to ensure that differences in the times of the output points do not introduce significant additional errors.

\begin{figure}
\includegraphics{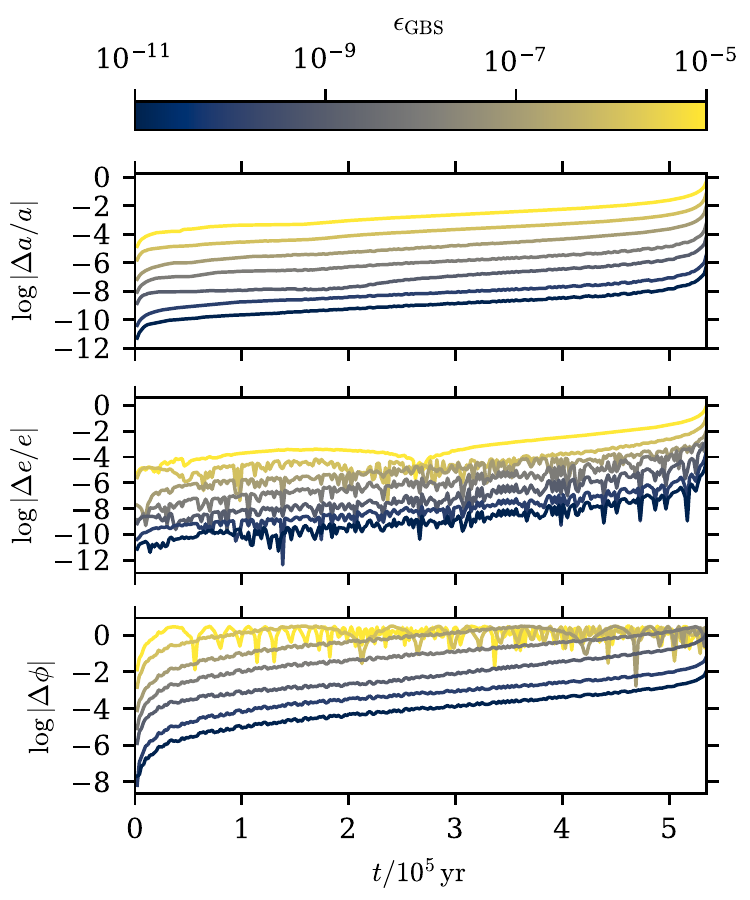}
\caption{
The relative error in the semimajor axis $a$ (top panel), the eccentricity $e$ (middle panel),
and the error in the orbital phase $\phi$ (bottom panel) between runs of an isolated SMBH binary
at different integrator tolerances $\epsilon_\mathrm{GBS}$.
The errors are calculated with respect to a reference run using a very high accuracy of $\epsilon_\mathrm{GBS}=10^{-12}$.
}
\label{fig:PN_tol_convergence}
\end{figure}

\begin{figure}
\includegraphics{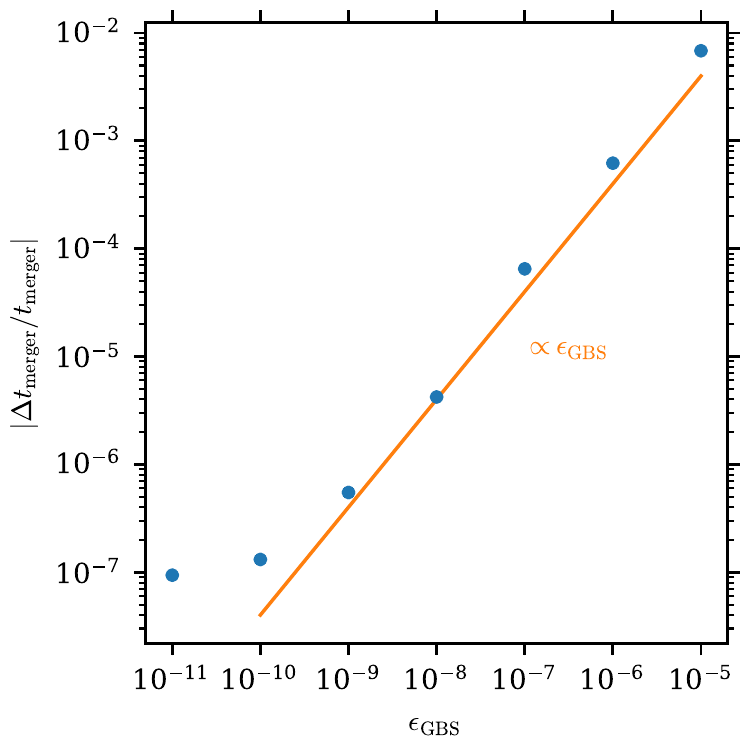}
\caption{
The relative error in the time when the SMBH binary system reaches the integrator
merger condition, as defined in Eq. \eqref{eq:merger_distance},
compared to a reference run using  $\epsilon_\mathrm{GBS}=10^{-12}$.
The error appears to scale nearly linearly with the tolerance, but saturates at the smallest tolerances where
the error is a fraction of an orbital period and is mainly due to the merger condition being
only evaluated after each full integrator step.
}
\label{fig:PN_tol_convergence_merger_time}
\end{figure}

The resulting relative errors in various orbital parameters
compared to a reference run using a tolerance of $\epsilon_\mathrm{GBS}=10^{-12}$
are shown in \autoref{fig:PN_tol_convergence}.
In addition, we show in \autoref{fig:PN_tol_convergence_merger_time} the relative error in the time of the binary
`merger' based on the merger condition defined in Eq. \eqref{eq:merger_distance}.
Larger tolerance values result in relatively large errors, however the errors converge well with
smaller tolerance values, so that tolerances below $\epsilon_\mathrm{GBS}\lesssim 10^{-7}$
match the reference solution to sub-percent level over the entire integration period.

However, following the orbital phase to within one full orbit
requires a relatively high tolerance of $\epsilon_\mathrm{GBS}\lesssim 10^{-9}$.
For longer integrations the errors accumulate to larger total values,
but since in typical simulations the effects of the surrounding stellar environment are significant and have
inherent inaccuracy due to the unphysically large particle masses,
tolerances of $\epsilon_\mathrm{GBS}=10^{-9}\textnormal{--}10^{-7}$ should provide the right balance between
capturing the SMBH binary orbital dynamics and the computational cost for most applications.

\subsection{Computational scaling}

The scalability of both the regularised MSTAR integrator \citep{2020MNRAS.492.4131R} and the \gadgetfour{} code
\citep{2021MNRAS.506.2871S} have been studied extensively in isolation.
How the combined \ketju{} code scales depends mainly on the relative computational cost
of the regularised integration compared to the rest of the system,
which depends on both the number of particles in the system and the details
of the particular simulation setup, which together determines how difficult the integration is to perform at
the set tolerances. Here we restrict our study to investigating the strong scaling
of the code, i.e. how quickly the same problem is solved when increasing the number of CPU cores.
We test the computational scaling for two cases where the regularised integration is either the dominant
contribution or only a small fraction of the total computational cost.

For the first test case, we run short simulations of an isolated spherical galaxy consisting of a $M_\star = 10^{11} \Msun$
stellar component with a $\gamma = 1.5$ Dehnen density profile and an effective radius of $R_\mathrm{e} = 5\,\mathrm{kpc}$
set in a DM halo with a mass of $M_\mathrm{DM} = 4\times10^{13}\, \Msun$ following a $\gamma = 1$ Dehnen profile.
Both the stellar and DM components are resolved using $10^7$ particles.
To create a system where the regularised integration is the dominant contribution,
we place a BH binary with masses of $M_{\bullet,1} = 10^8 \ \Msun$
and $M_{\bullet,2}=3.33 \times 10^8 \ \Msun$ on an orbit with
a semimajor axis of $a = 3\,\mathrm{pc}$ and an eccentricity of $e = 0.6$
in the centre of the galaxy.
We further set the  stellar softening length to
$\epsilon = 7\,\mathrm{pc}$ and $\rreg = 21\,\mathrm{pc}$,
yielding about 4000 particles in the region and thus making
regularised integration computationally dominant over the softened dynamics
integration in \gadgetfour{}.

For the second test case where the regularised integration is only a small fraction of the computational cost,
we use a single BH of mass $M_\bullet = 10^8 \ \Msun$ in the centre of the galaxy
with a stellar softening length of $\epsilon = 2\,\mathrm{pc}$
and a small regularised region radius of $\rreg = 6\,\mathrm{pc}$,
yielding about 600 particles in the region.
In typical applications, the main factor in deciding the size of the regularised
region is the lower limit imposed by the chosen softening length.
The softening lengths used here are chosen to match what would be
typically adopted for the chosen region sizes,
as the softening length affects the performance of the main \gadgetfour{} integration.
We run these systems for $10^5 \,\mathrm{yr}$ using varying numbers of CPU cores
ranging from $N=16$ to $1024$.

\begin{figure}
\includegraphics{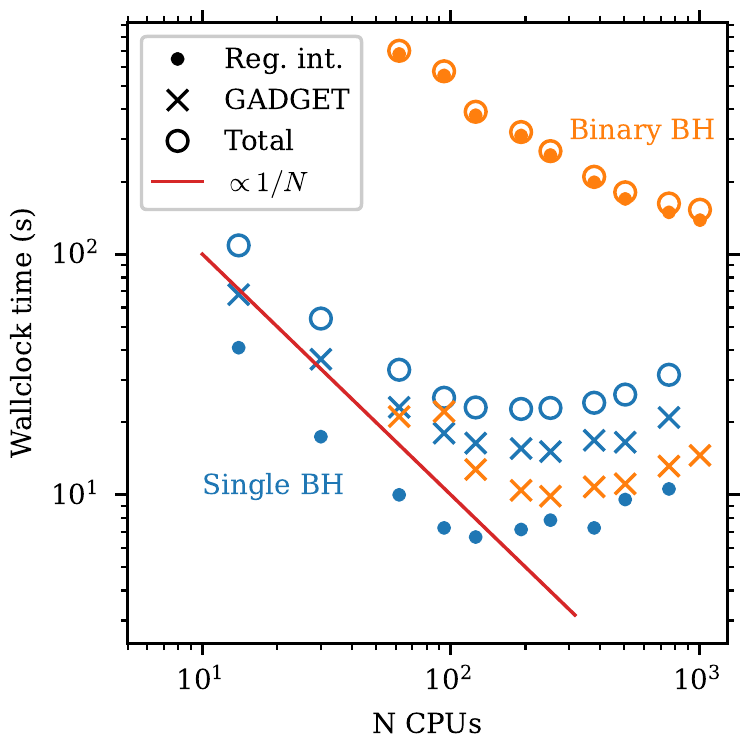}
\caption{
The strong scaling of the code in two different systems with a total of $10^7$
stellar particles.
In the first test case the regularised region contains $\sim 4000$
particles and an SMBH binary (orange symbols), while in the second test case the system includes only a single SMBH surrounded by a
regularised region containing $\sim 600$ particles (blue symbols).
The figure shows the wallclock times taken by the regularised integration (solid circles), the \gadgetfour{} part (cross symbols) and
the combined total \ketju{} integration time (open circles), when the simulations are run using different numbers $N$ of CPUs.
The ideal scaling slope of $\propto 1/N$ is also indicated for reference.
}
\label{fig:strong_scaling}
\end{figure}

The results of this strong scaling test are shown in \autoref{fig:strong_scaling}.
The total time taken by the code can be seen to scale nearly ideally at low CPU counts in both systems,
whereas at very high core counts in particular the single BH case begins to even slow down.
For both cases the time taken in the \gadgetfour{} part is very similar,
although slightly less wallclock time is spent for the binary BH system for which the softening length is somewhat larger.
The scaling of the \gadgetfour{} part of the calculations stops at around a 100 CPUs for this system,
corresponding to the size of a single node on the Mahti supercomputer.

The scaling of the regularised integration also stops around the same point for the
single BH system, so that for this system the code cannot make use of a larger number of CPUs.
On the other hand, in the binary BH case the simulation time is dominated by the regularised integration,
as the region is more computationally expensive due to both the larger particle count
as well as the presence of the BH binary.
The higher particle count allows the regularised integrator to scale nearly ideally
to a much higher number of CPUs, allowing also the total run time to scale well
despite the fact that the \gadgetfour{} part is also becoming slightly slower for
larger CPU numbers in this system.

These results show how the scalability of the code depends on the relative cost
of the regularised integration, and how in cases where the regularised integration is dominant
the code can make use of much larger CPU counts than would be sensible for a standard \gadgetfour{}
simulation of a system of the same size.
These results are also in line with the extensive scaling test of the regularised integrator
presented in \citet{2020MNRAS.492.4131R}, where it was found that the integrator
scales well until $N_\mathrm{part} \sim 10\, N_\mathrm{CPU}$.
To avoid the slowing down of the integration at excessive core counts,
the code automatically limits the number of CPUs
allocated to each region based on its particle count.

\section{Formation of cores in multiple mergers}
\label{Sec:Formation_cores}

\subsection{Simulations}

\begin{table}
\caption{
The stellar mass, effective radius $R_\mathrm{e}$, and the scale radii $r_\mathrm{s}$
for the Hernquist sphere stellar and DM components
of the individual initial galaxies in the runs with different numbers $N_\mathrm{gal}$ of galaxies.
}
\label{table:gal_scale_radii}
\centering
\begin{tabular}{ccccc}
\hline
$N_\mathrm{gal}$ & $M_\star/(10^{11}\,\Msun)$  & $R_\mathrm{e}/\mathrm{kpc}$ & $r_\mathrm{s,\star}/\mathrm{kpc}$ & $r_\mathrm{s,DM}/\mathrm{kpc}$\\
\hline
2  & 2.00 & 4.79 & 2.65  & 233\\
3  & 1.33 & 3.76 & 2.08  & 182\\
5  & 0.80 & 2.77 & 1.53  & 134\\
\hline
\end{tabular}
\end{table}

As an example of the types of problems that can be studied with the public version
of \ketju{} presented here, we perform a simplified study of how the number of progenitor
galaxies affects the size of the core in the final merger remnant.
The formation of cores in massive elliptical galaxies, which is commonly attributed to core scouring
by SMBH binaries, has been already studied before using an earlier version of \ketju{} \citep{2018ApJ...864..113R},
and also to a great extent by various other authors with different codes \citep[e.g.][]{2001ApJ...563...34M,2002MNRAS.331L..51M,2006ApJ...648..976M,2012ApJ...749..147K,2021MNRAS.502.4794N,2021ApJ...922...40D}.

Here the main difference to most other studies of this process is that multiple mergers are allowed to occur in quick succession
without necessarily allowing each SMBH binary to merge first before the next galaxy merger.
These simulations thus allow us to demonstrate how \ketju{} can be used to capture the complex
dynamics in systems with multiple interacting SMBHs.

We simulate the mergers of three different systems with two, three, and five
identical galaxies in the initial conditions, respectively.
All systems are set to have the same total stellar mass of $M_{\star}=4\times 10^{11} \ \Msun$,
a total DM mass of $M_\mathrm{DM}=7\times 10^{13} \,\Msun$, and a total black hole mass
of $M_{\bullet}=2.5\times 10^9 \ \Msun$, with the total masses divided either between two, three or five
galaxies, depending on the simulation.

The DM particles have a mass of $m_\mathrm{DM}=5\times 10^6 \,\Msun$ and a softening length of $\epsilon_{\rm DM}=100\,\mathrm{pc}$,
while the stellar particles have a mass of $m_{\star}=10^5 \,\Msun$ and a softening length of $\epsilon_{\star}=15\,\mathrm{pc}$.
The BHs are surrounded by regularised regions with a radius of $r_{\rm region}=45\,\mathrm{pc}$,
with all BH-BH and BH-star interactions being non-softened.
Only stellar particles are allowed to enter the regularised regions, while the DM acts as a smooth perturbing
background on the dynamics and has softened interactions with all particles.
With these parameters, the mass ratio between the initial SMBHs and stellar particles is
$m_{\bullet}/m_{\star}\geq 5000$, and the final merged galaxy has about four million stellar particles,
which should ensure converged behaviour of the SMBH binary hardening based on the
tests presented in Section~\ref{sec:binary_hardening}.

The initial galaxies are modelled as multicomponent Hernquist spheres, setup using Eddington's
formula \citep{BTgaldynbook} and the methods described in \citet{2018ApJ...864..113R}.
The projected stellar half-mass radii are set based on the galaxy stellar mass $M_\star$ by
\begin{equation}
R_\mathrm{e} = 3.16 \left(\frac{M_\star}{10^{11}\Msun}\right)^{0.6}\,\mathrm{kpc},
\end{equation}
which approximately matches the observed size-stellar mass relation \citep[table 1]{2012ApJ...746..162N}.
The DM scale radius is then set so that the central DM fraction is $f_\mathrm{DM}=0.2$.
The scale radii of the initial galaxies are listed in \autoref{table:gal_scale_radii}.
Each galaxy is also given an SMBH at the centre, with initially zero spin for simplicity.

The galaxies are placed on initial orbits that result in rapid galaxy mergers, with
relatively small initial separations compared to the sizes of their DM haloes. This is done
in order to reduce the simulation time, and should not significantly affect the simulation
outcome, as the stellar components of the galaxies do not overlap.
The two-galaxy system is placed on a nearly parabolic orbit at an initial separation
of $30\,\mathrm{kpc}$,
while the three-galaxy system has two galaxies on a similar orbit with an initial
separation of $45\,\mathrm{kpc}$ with the third galaxy at rest at a distance of $35\,\mathrm{kpc}$
from the midpoint of these two galaxies.
The five-galaxy system is initialised by generating random positions for the galaxies
within a box with $80\,\mathrm{kpc}$ on a side, and then generating normally distributed
velocities so that the system is in virial equilibrium, assuming that the galaxies
are point masses.

To demonstrate the cosmological integration capabilities of \ketju{}, we place the
two- and three-galaxy systems in the same periodic box with a $10 \,h^{-1} \mathrm{Mpc}$ comoving side length,
ensuring that the separation of the systems is large enough not to interact with each other.
We simulated this system starting from redshift $z=1$ using a cosmology with
$\Omega_\mathrm{m} = 0.3$, $\Omega_\Lambda=0.7$, and $h = H_0/100\,\mathrm{km\,s^{-1}\,Mpc^{-1}} = 0.7$.
These choices do not meaningfully affect the dynamics of the merging systems,
but simply serve to demonstrate that the code operates correctly also in a cosmological comoving setting.
We run the five-galaxy system in standard non-comoving coordinates and non-periodic space.
For the analysis of the results from both simulations we will use physical coordinates,
and set time $t=0$ at the start of the simulation.
In order to study the effect of the SMBHs on the merger remnant structure,
we also run versions of the simulations where the initial galaxy models are generated
without central SMBHs.

For simplicity, we disable the GW recoil kicks during SMBH mergers.
For the initial non-spinning SMBHs with equal masses such kicks would vanish in
any case, but for the second- and third-generation mergers that are possible
in the systems with multiple merging galaxies they might lead to the ejection
of the final remnant SMBH from the galaxy \citep{2022ApJ...929..167M}.
As the kick magnitudes sensitively depend on the SMBH spins and are effectively random,
including them would add an unnecessary layer of complexity for interpreting the
results of these simplified demonstration simulations.
However, we note that the recoil kicks can significantly contribute
to the formation of cores \citep{2021MNRAS.502.4794N}, and they should thus
be included in more comprehensive studies of the core formation mechanism.

\subsection{SMBH dynamics}

\begin{figure}
\includegraphics{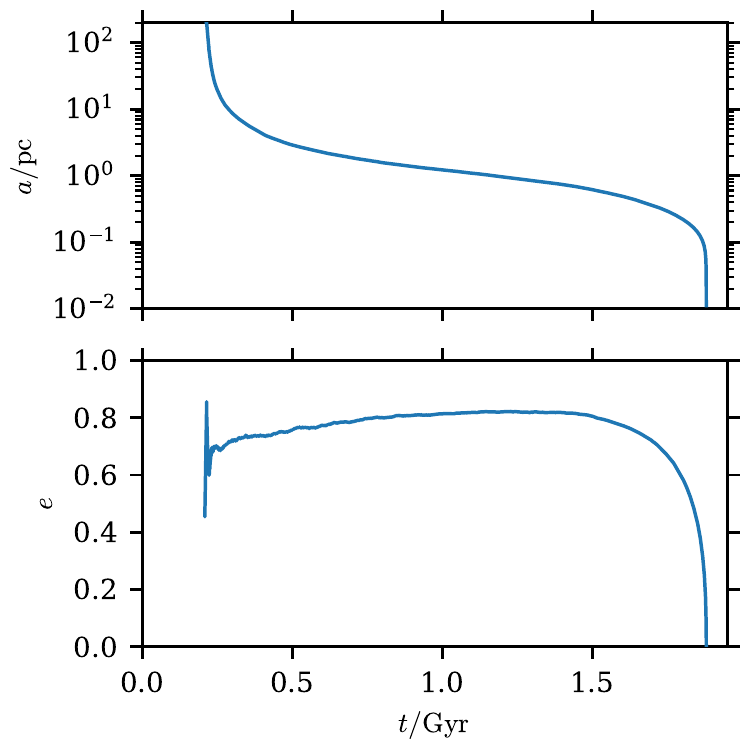}
\caption{
The semimajor axis $a$ (top) and eccentricity $e$ (bottom) of the SMBH binary formed in the two-galaxy merger.
The initial binary eccentricity is $e\sim 0.7$ and the binary hardens due to stellar scattering with the
final GW emission driven merger occurring $\sim 1.5 \ \rm Gyr$ after its formation.}
\label{fig:binary_merger_bh_pars}
\end{figure}

\begin{figure}
\includegraphics{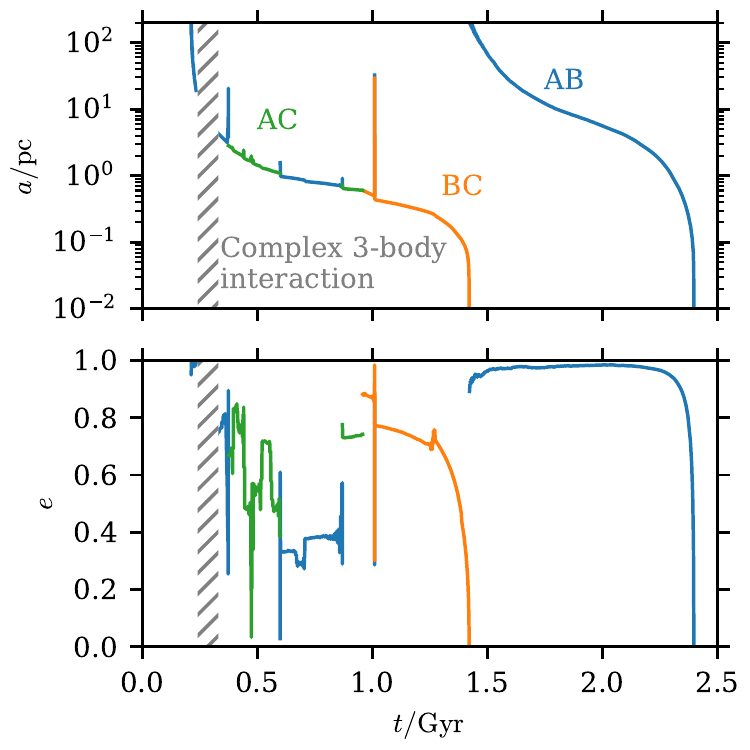}
\caption{
The semimajor axis $a$ (top) and eccentricity $e$ (bottom) of the SMBH binaries formed in the three-galaxy merger.
In the hatched region the three SMBHs in the system have strong interactions with each
other, and the orbital parameters cannot be meaningfully calculated.
For most of the its evolution, the SMBHs are in a hierarchical triplet configuration
with a very wide outer orbit, with occasional close interactions that can change the component
SMBHs in the inner binary, as shown in the figure.
The outer orbit parameters are not shown for simplicity, as most of the time they
are significantly affected by the stellar mass inside the very wide outer orbit.
}
\label{fig:triplet_merger_bh_pars}
\end{figure}

\begin{figure}
\includegraphics{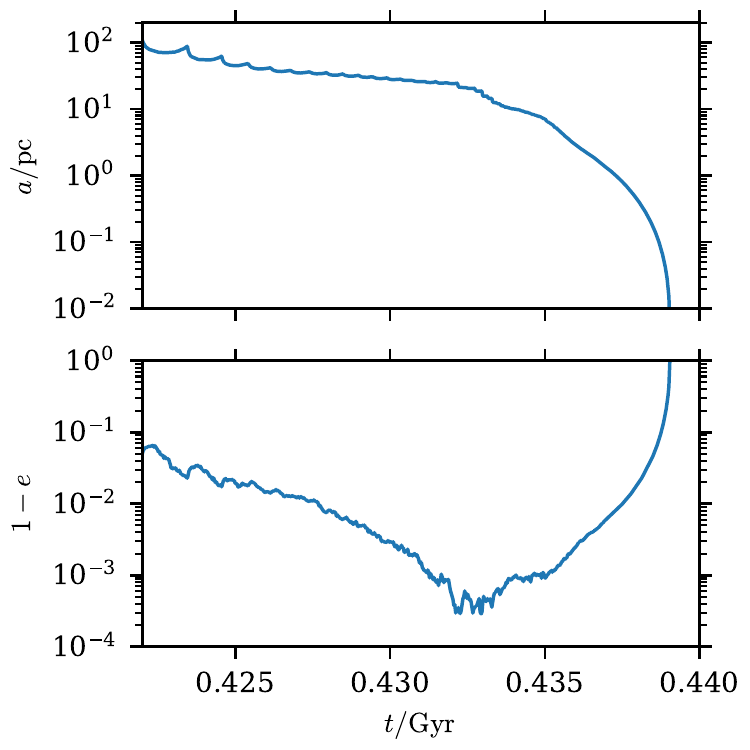}
\caption{
The semimajor axis $a$ (top) and eccentricity $e$ (bottom) of the binary formed by SMBHs C and D in the five-galaxy merger.
The eccentricity is shown using $1-e$ on a  log-scale to make the extremely high eccentricity
before the rapid GW-driven merger visible.
}
\label{fig:group_merger_bh_pars_CD}
\end{figure}

\begin{figure*}
\includegraphics{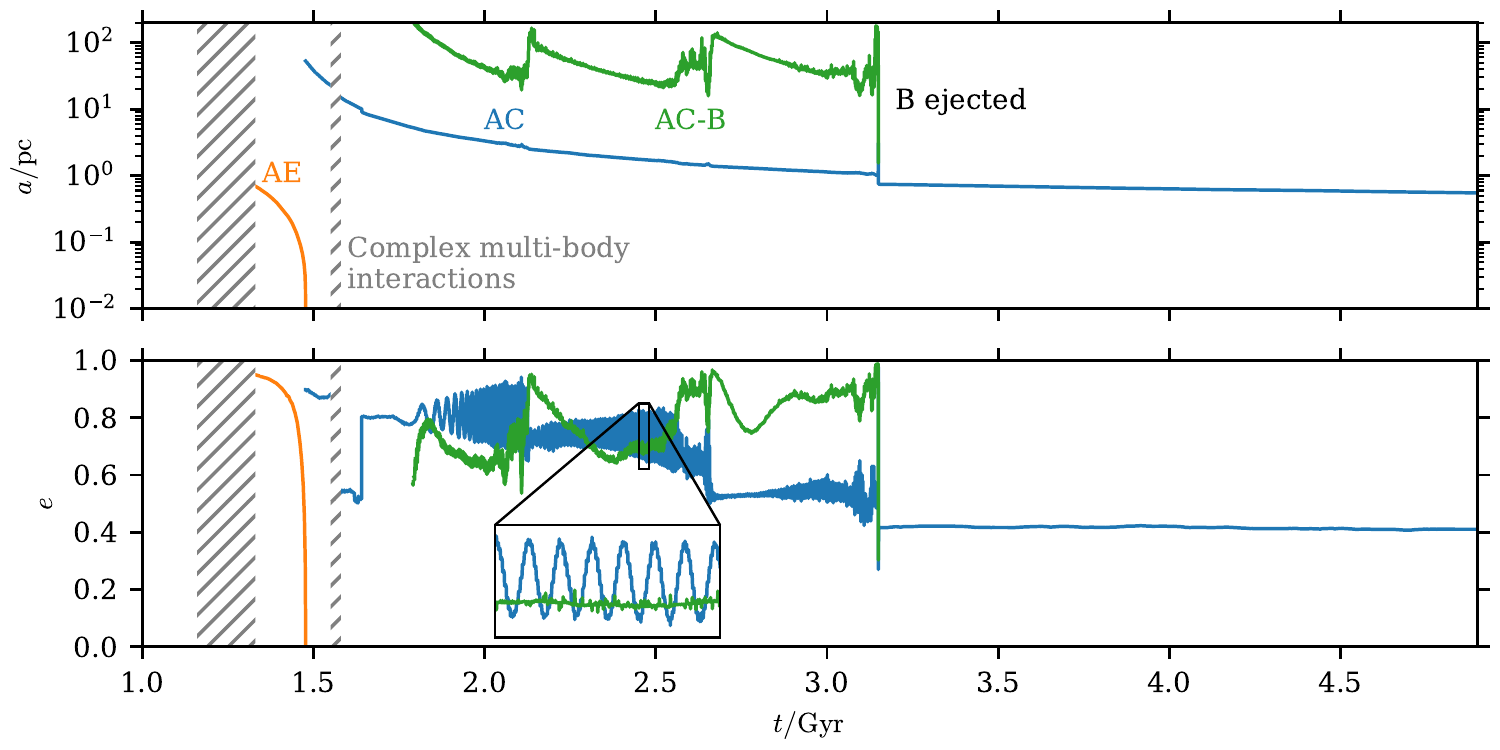}
\caption{
The semimajor axis $a$ (top) and eccentricity $e$ (bottom) of the SMBH binaries and hierarchical triplet
that form after the CD binary has merged.
In the hatched regions the multiple SMBHs in the system have strong interactions with each
other, and the orbital parameters cannot be meaningfully calculated.
The AC-B parameters show the orbit of SMBH B around the AC binary in a hierarchical triplet configuration.
SMBH B has multiple close interactions with the binary that throw it onto a wider orbit,
before finally being ejected completely from the galaxy.
The inner AC binary also undergoes von Zeipel--Lidov--Kozai oscillations during this hierarchical triple phase.
}
\label{fig:group_merger_bh_pars_main}
\end{figure*}

The behaviour of the SMBHs in the two-galaxy merger is relatively simple,
as can be seen from the orbital parameters in \autoref{fig:binary_merger_bh_pars}.
The SMBHs form a binary on a moderate-eccentricity $(e\approx 0.7)$ orbit in the centre of the merged
galaxy after about $200\,\mathrm{Myr}$.
This binary then hardens due to stellar scattering and finally merges due to GW emission
about $1.5\,\mathrm{Gyr}$ after it formed.

In the three-galaxy merger the behaviour is more complex, as can be expected since the
chosen initial condition results in all three galaxies merging almost simultaneously.
The orbital parameters of the bound binaries formed in this system are shown
in \autoref{fig:triplet_merger_bh_pars},
with the SMBHs labelled A, B and C based on the internal simulation particle IDs and the
SMBH binaries labelled with the corresponding letter pairs.
Most of the time the system is in a hierarchical triplet state with a fairly wide
outer orbit, but occasionally there are strong interactions between all three SMBHs
when the outer orbit shrinks due to dynamical friction and stellar scattering.
During these strong interactions the components of the central binary are sometimes
exchanged, so that while A and B are the first SMBHs to form a bound binary,
it is actually B and C that merge first due to GW emission.
The time until the first SMBH merger from the formation of the first binary is slightly
shorter here than in the two-galaxy merger, partly due to the
strong interactions between the SMBHs contributing to the hardening of the inner binary.
The remaining AB binary is left on a wide but highly eccentric orbit $(e\approx 0.98)$, and it hardens
efficiently enough to merge roughly $\sim 1 \,\mathrm{Gyr}$ later.

In the five-galaxy merger we label the SMBHs as A--E in order of their internal simulation particle IDs.
After about $400\,\mathrm{Myr}$, the galaxies hosting SMBHs C and D have merged, and the SMBHs form a bound binary.
This binary quickly evolves to have an extremely high eccentricity ($e\approx 0.9995$) and then merges due to strong
GW emission within only a few million years.
The remnant SMBH is still labelled as C after this merger.
The evolution of the orbital parameters of this binary is shown in \autoref{fig:group_merger_bh_pars_CD}.

Over the next few hundred million years, the remaining galaxies merge together and the SMBHs
begin interacting in the centre of the merged galaxy remnant.
The evolution of the orbital parameters of the bound systems that are formed during these interactions
is shown in \autoref{fig:group_merger_bh_pars_main}.
First, after a period of complex interactions between all four remaining SMBHs in the system,
SMBHs A and E form a binary on a high-eccentricity orbit  $(e\approx 0.95)$ and sub-parsec separation.
This binary is already in the GW emission dominated regime of its orbital evolution
and it is driven to merge within about $150\,\mathrm{Myr}$.

Rapidly after this merger, SMBHs A and C form a wider binary, which interacts with
SMBH B throwing it out to a wide orbit.
These three SMBHs form a hierarchical triplet where the inner AC binary undergoes
von Zeipel--Lidov--Kozai oscillations \citep{1962P&SS....9..719L}.
As the orbit of SMBH B shrinks due to interactions with stars,
it has several
strong interactions with the central binary that throw it back out to a wide orbit,
and finally ejects it completely from the galaxy.
After this, the hardening of the AC-binary due to stellar scattering is quite weak.
Based on  analytical results for the stellar scattering driven hardening \citep{1996NewA....1...35Q}
and GW emission \citep{1964PhRv..136.1224P},
we estimate that the binary would merge only at $t\sim 11\,\mathrm{Gyr}$.
Running the simulation for this long would not bring enough additional benefit compared to the computational cost,
and we therefore stop the simulation already at $t\approx 5\,\mathrm{Gyr}$.

\subsection{Remnant density profiles and core sizes}

\begin{figure}
\includegraphics{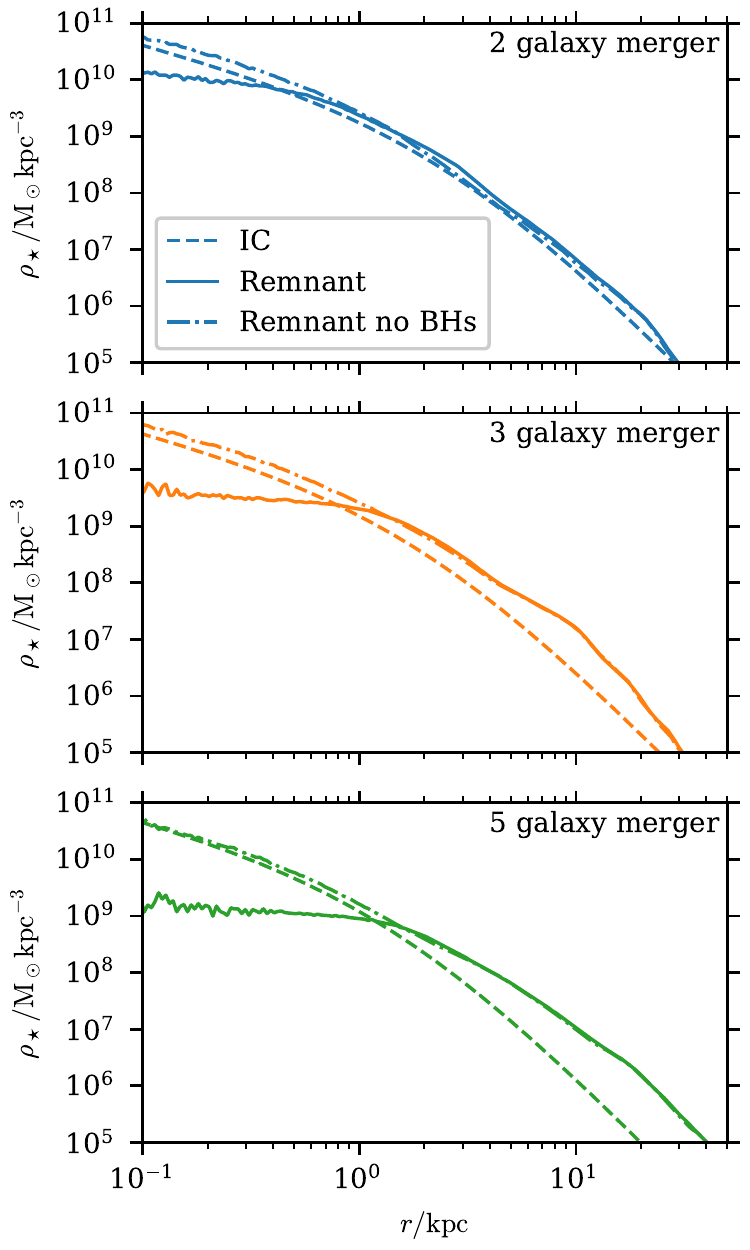}
\caption{
Stellar density profiles of the initial condition (IC) galaxy models,
and the final merger remnants of the merging systems with different numbers of galaxies. In the
simulations without BHs the initial density profile is retained, as expected.
}
\label{fig:demo_runs_density_profiles}
\end{figure}

\begin{figure}
\includegraphics{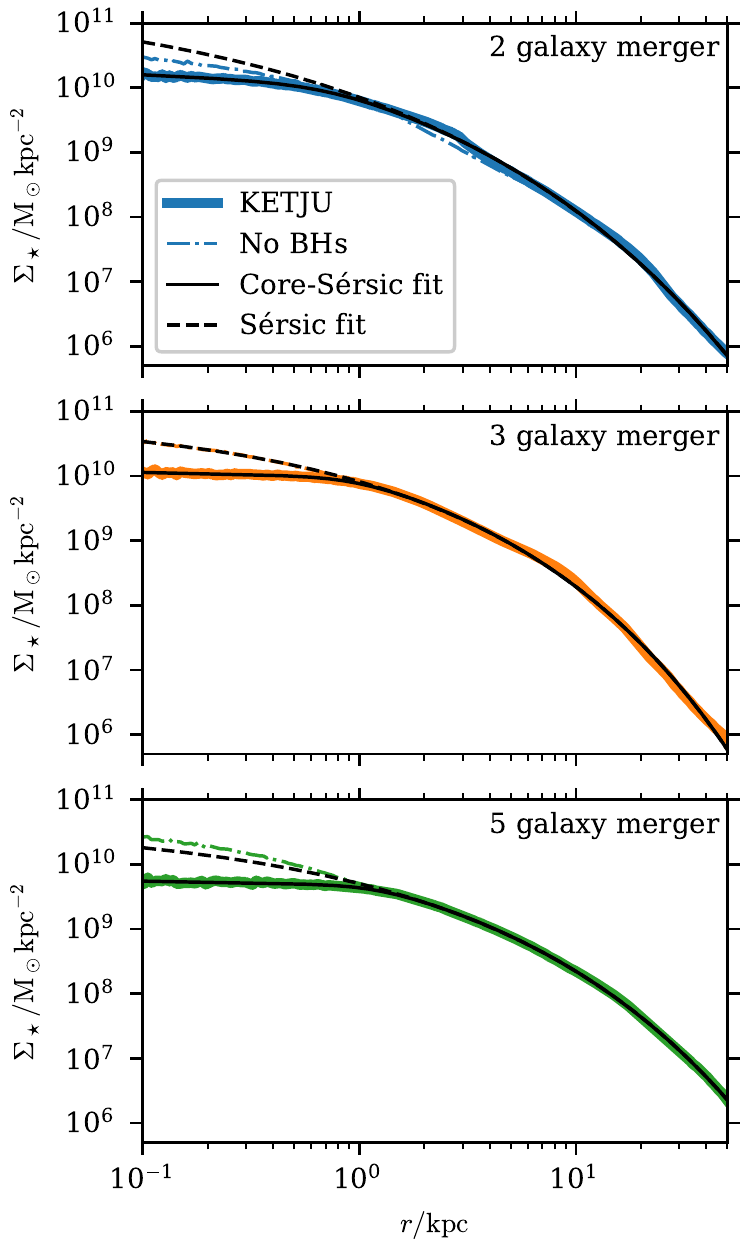}
\caption{
Projected stellar density profiles of the merger remnants as in \autoref{fig:demo_runs_density_profiles},
the fitted core-S\'ersic profiles, and the S\'ersic profiles
equivalent to the outer region of the core-S\'ersic fits.
}
\label{fig:demo_runs_projected_profiles}
\end{figure}

\begin{table*}
\centering
\caption{
Core-S\'ersic profile fit parameters and calculated mass deficits $M_\mathrm{def}$
for the merger remnant galaxies with different numbers $N_\mathrm{gal}$ of progenitors.
The profile parameters are  the break radius $r_\mathrm{b}$, the effective radius $R_\mathrm{e}$,
the inner power law slope $\gamma$, the S\'ersic index $n$, and the transition parameter $\alpha$.
Mass deficit results are shown for two different methods for calculating the mass deficit, see the text for details.
Note that the final SMBH mass in the galaxy $M_\bullet$ differs between the runs due to mass loss to GW emission
and the ejection of an SMBH in the five-galaxy run.
}
\label{tab:demo_runs_fit_params}
\begin{tabular}{ccccccccccc}
\hline
$N_\mathrm{gal}$ & $r_\mathrm{b}/\mathrm{kpc}$ & $R_\mathrm{e}/\mathrm{kpc}$ & $\gamma$ & $n$ & $\alpha$ & $M_\mathrm{def,1}/10^9 \Msun$ & $M_\mathrm{def,1}/M_\bullet$ & $M_\mathrm{def,2}/10^9 \Msun$ & $M_\mathrm{def,2}/M_\bullet$ & $M_\bullet/10^9 \Msun$\\
\hline
2 & 0.63 & 4.37 & 0.169 & 3.2 & 2.8 & 11.1 & 4.67 & 7.4 & 3.13 & 2.38\\
3 & 0.85 & 4.74 & 0.074 & 2.4 & 4.1 & 11.5 & 4.90 & 14.0 & 5.99 & 2.34\\
5 & 1.09 & 7.84 & 0.060 & 2.6 & 4.3 & 9.0 & 4.71 & 14.0 & 7.36 & 1.90\\
\hline
\end{tabular}
\end{table*}

The density profiles of the progenitor and merger remnant galaxies in the simulations are shown
in \autoref{fig:demo_runs_density_profiles}.
The remnant profiles in the \ketju{} runs with SMBHs show clearly flattened central cores,
with the size of the core increasing with the number of progenitor galaxies.
In the runs without SMBHs the merger remnants retain the steepness of the progenitor density profiles
also in the central parts \citep{2004MNRAS.349.1117B}. This shows that the
presence of the SMBHs is responsible for the formation of these cores, in line with the findings of
previous studies \citep[e.g.][]{2001ApJ...563...34M,2006ApJ...648..976M,2018ApJ...864..113R}.

To quantify the size of the cores in the merger remnants, we fit core-S\'ersic
profiles \citep{2003AJ....125.2951G}
\begin{equation}
\Sigma(r) = \Sigma' \Bigl[1+\Bigl(\frac{r_\mathrm{b}}{r} \Bigr)^\alpha \Bigr]^{\gamma/\alpha}
            \exp\Biggl[-b_n \Biggl( \frac{r^\alpha + r_\mathrm{b}^\alpha}{R_\mathrm{e}^\alpha} \Biggr)^{1/(\alpha n)} \Biggr]
\end{equation}
to the projected mass density profiles of the galaxies.
Here $n$ is the S\'ersic index, $R_\mathrm{e}$ is the effective radius, $r_\mathrm{b}$ is the break radius,
$\gamma$ is the inner power law slope, $\alpha$ is a parameter controlling how sharp the transition between the power law and S\'ersic profiles is,  and $b_n \approx 1.9992n - 0.3271$.
The fits were performed in the range $0.1 < r/\mathrm{kpc} < 50$.
The fitted parameters are shown in \autoref{tab:demo_runs_fit_params},
and the corresponding profiles are shown in \autoref{fig:demo_runs_projected_profiles}.

The fitting results confirm what can also be seen visually from the density profiles:
the size of the core, quantified by the break radius, increases for increasing numbers of progenitor galaxies,
with the five-galaxy merger having a core that is almost twice as large as the core in the binary merger.
The density profile within the core region also becomes flatter when the number of progenitors
increases, and the overall size of the galaxy, as defined by the effective radius, also grows.
These are both natural consequences of the stellar mass being displaced from the central
regions due to interactions with the SMBHs.

The derived core size for our two-galaxy merger agrees well with the results of the binary merger of two Hernquist spheres $(\rho \propto r^{-1})$ with a similar final BH mass, obtained in \citet{2018ApJ...864..113R}. The fits also produce S\'ersic indices of the merger remnants that are somewhat on the low-side, $2.4\leq n \leq 3.2$, compared to $n\gtrsim 4$ typically observed for elliptical galaxies of similar masses. However, this is a well-known feature of the Hernquist model \citep{2006MNRAS.369..625N} and also in good agreement with the findings of \citet{2018ApJ...864..113R}.

When comparing the simulated core sizes to observations, as defined by the observed $M_\bullet \textnormal{--} r_\mathrm{b}$ relation \citep{2016Natur.532..340T,2021ApJ...908..134D}, we find that the mean prediction is $r_\mathrm{b} \sim 200\,\mathrm{pc}$ for the SMBH mass in these systems, albeit with a rather large scatter. The core sizes are rather sensitive to the initial density profile
of the galaxies, and as shown in \citet{2018ApJ...864..113R} steeper density profiles $(\rho \propto r^{-1.5})$ result in core sizes
that are typically a factor of $\sim 2 \textnormal{--} 3$ times smaller than the rather flat Hernquist profile. Steeper density profiles provide a better match to real observed elliptical galaxies, but are also more expensive to simulate, due to the higher central stellar density. Thus, as the main goal of this paper is to demonstrate the operation of the \ketju{} code in various simulation settings, we opted to run our simulations with the less steep Hernquist profiles.

\subsection{Mass deficits}

The size of the cores formed in mergers of galaxies with SMBHs can also be quantified
in terms of the mass deficit, defined as the difference in stellar mass contained in the core region
when compared to an equivalent galaxy without a core.
The mass deficit is expected to scale with the total SMBH mass,
with e.g. \citet{2006ApJ...648..976M} proposing a simple scaling based on the number of mergers $N_\mathrm{merger}$,
\begin{equation} \label{eq:merritt_mdef}
M_\mathrm{def} \approx 0.5 N_\mathrm{merger} M_\bullet,
\end{equation}
whereas other studies have also found larger values, with the results depending to a significant degree also
on the exact methods used to define the mass deficit \citep{2021ApJ...922...40D}.
Observations show that the relative mass deficit is $M_\mathrm{def}/M_\bullet$ in the range $0.5$--$10$ \citep[e.g.][]{2013AJ....146..160R}.

To calculate this mass deficit for our runs, we use two different approaches.
In the first method we calculate the deficit $(M_\mathrm{def,1})$ based on the difference between the fitted
core-S\'ersic profile and a S\'ersic profile fitted to the density profile outside the core region.
In the second approach $(M_\mathrm{def,2})$, we use as a reference the equivalent galaxies
and merger remnants from the corresponding runs without SMBHs, and calculate the difference in the mass contained within
the region where the reference galaxy has a higher stellar density.
The results of this calculation are listed in \autoref{tab:demo_runs_fit_params},
and it can be seen that the two different approaches give reasonably similar results.

In all cases the ratio $M_\mathrm{def}/M_\bullet$ is larger than predicted by
the relation \eqref{eq:merritt_mdef}, but consistent with the range of observed
mass deficits \citep[e.g.][]{2013AJ....146..160R,2021ApJ...908..134D}.
It is also interesting to note that using the second definition,
the mass deficit is very similar in both the three-
and five-galaxy mergers, and clearly larger than in the two-galaxy merger.
The complex interactions in the multi-SMBH systems found in these systems might
contribute to the relatively large mass deficits, as the repeated ejection of an SMBH to wide orbits within
the galaxy could increase the effect it has on the stellar system.
\citet{2012MNRAS.422.1306K} also found similarly large mass deficits in systems with multiple interacting SMBHs.
On the other hand, using the first definition for the mass deficit there does not appear
to be any meaningful difference between the different mergers.
As is the case with the core radii, the large mass deficit values are likely mainly due to the simplified Hernquist profile
initial conditions, with e.g.\ \citet{2021ApJ...922...40D} finding a factor of $\sim 3$ larger mass deficts for $\rho \propto r^{-1}$
profiles compared to steeper initial density profiles with $\rho \propto r^{-1.5}$.

\subsection{Velocity anisotropy profiles}

\begin{figure}
\includegraphics{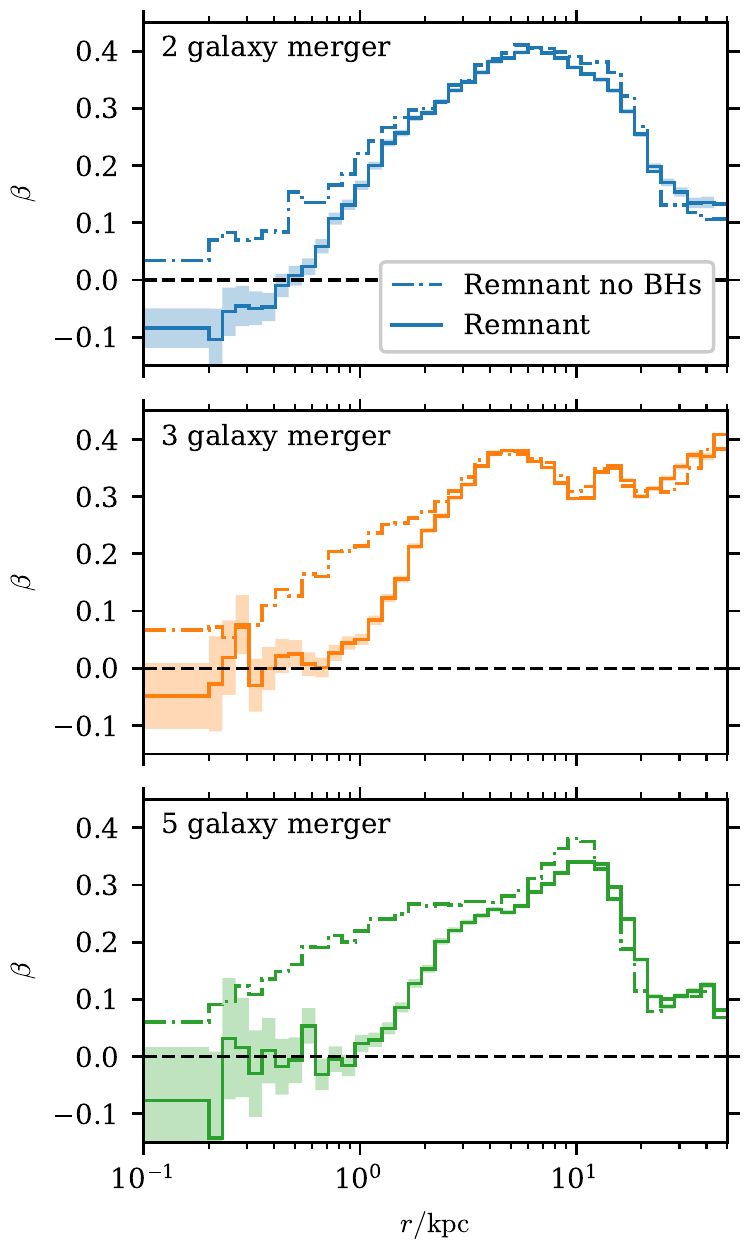}
\caption{The stellar velocity anisotropy $\beta(r)$ profiles of the final merger
  remnants of the merging systems with different numbers of galaxies. The isotropic progenitor
  galaxies are shown as the black horizontal dashed line with $\beta(r)=0$.
  The $\beta(r)$ profiles are computed in logarithmically spaced bins, with the shaded regions showing the
  estimated $1\sigma$-uncertainty due to the finite number of particles in each radial bin.
  All profiles show the expected behaviour of an increasingly more tangentially biased
  velocity distribution at small radii due to the influence of the SMBH binary.}
\label{fig:demo_runs_beta_profiles}
\end{figure}

The dynamical interaction between the SMBHs and the stellar particles is expected
to lead to the preferential ejection of particles on radially biased orbits,
and result in a tangential bias in the orbits of the stellar particles near the
centre of the final merger remnant \citep{1997NewA....2..533Q,2001ApJ...563...34M,2014ApJ...782...39T}.

This effect can be quantified using the velocity anisotropy  parameter $\beta$,
which is defined as \citep[e.g.][]{BTgaldynbook}
\begin{equation}
\beta = 1 - \frac{\sigma_\theta^2 + \sigma_\phi^2}{2 \sigma_r^2},
\end{equation}
where $\sigma_r$, $\sigma_\theta$, and $\sigma_\phi$ are the stellar velocity
dispersions along the spherical coordinate directions.
In \autoref{fig:demo_runs_beta_profiles} we show the radial $\beta(r)$-profiles of the
merger remnants of the three runs, with two, three and five galaxy progenitors (top to bottom).
The velocity anisotropy is computed in radial bins with logarithmic spacing around the
centre of the final stellar component, and the shaded regions show the
estimated $1\sigma$-uncertainty due to the finite number of particles in each radial bin.

The profiles clearly show the expected behaviour of a slight tangential bias ($\beta(r)<0$)
in the centre that transitions to radial bias ($\beta(r)>0$) in the outer parts of the galaxy, which
is typical for collisionless mergers \citep{2021MNRAS.508.4610F}. The effect of the SMBHs is also apparent when comparing
the profiles of the runs with (solid lines) and without SMBHs (dash-dotted lines), as the runs without SMBHs show
consistently larger values of $\beta(r)$. Finally, we note that the profile of the two-galaxy merger is also very similar to
the profiles published by \citet{2018ApJ...864..113R} for binary mergers with similar SMBH masses.

\section{Summary and conclusions}
\label{sec:Conclusions}

In this paper we have presented the new publicly available version of the \ketju{} code, as implemented in \gadgetfour{}.
We have demonstrated how the code can follow the dynamics of SMBHs to spatial scales significantly below the
softening length used for stellar and other simulation particles,
which is not possible with standard \gadgetfour{} or other similar softened $N$-body codes.

The presented code tests show that dynamical friction acting on SMBHs sinking in a stellar systems
is captured well even at relatively low particle resolutions. As all BH-BH and BH-star interactions are non-softened
in \ketju{}, a comparable \gadgetfour{} run would have to be run with much smaller softening lengths and at a significantly
higher computational expense (see \autoref{fig:BH_plummer_sinking}).
We also show that the hardening rate of a SMBH binary due to stellar
scattering converges when the mass ratio between the BH and stellar particles is $M_{\bullet}/m_{\star}\gtrsim 10^3$. For the applications presented here
this corresponds to galaxy models with $\gtrsim 1$ million stellar particles. The convergence of the hardening rate can be explained by the reduced impact
of the Brownian motion of the SMBH binary at higher mass resolutions, i.e. the stellar mass entering and exiting the loss cone varies more smoothly with increased
mass resolution. The convergent behaviour of the hardening rate is similar to
what has been previously found in studies using other high-accuracy $N$-body codes \citep{2006ApJ...642L..21B,2011ApJ...732...89K,2017MNRAS.464.2301G}.
In line with this finding, we also confirmed that \gadgetfour{} does not suffer from the type of errors in modelling the loss cone refilling that have been previously reported for an earlier version of the GADGET code \citep{2017MNRAS.464.2301G}.

While the interactions between BHs and stars are always non-softened in \ketju{}, the star-star interactions can also be optionally softened even if the
stars reside within the regularised region. Including this softening length will reduce energy errors
as particles are moving in and out of the region and it is particularly important in simulations where a large stellar mass,
comparable to the black hole mass, is contained within the regularised region.

The accuracy of the regularised MSTAR \citep{2020MNRAS.492.4131R} integrator,
which uses a Gragg--Bulirsch--Stoer extrapolation scheme is controlled by a per-step relative error tolerance parameter, $\epsilon_\mathrm{GBS}$. In addition,
due to the time transformation used in algorithmic regularisation an output time relative tolerance parameter $\epsilon_t$ is used to control the accuracy
of the iterative solution. Tests of varying $\epsilon_\mathrm{GBS}$ for a fixed value of $\epsilon_t=10^{-10}$ showed that values of $\epsilon_\mathrm{GBS}\lesssim 10^{-7}$
match the reference solution run at $\epsilon_\mathrm{GBS}=10^{-12}$ to a sub-percent level for a typical SMBH binary system. However, following the orbital phase to within one full orbit requires higher precision of  $\epsilon_\mathrm{GBS}\lesssim 10^{-9}$, which is thus required if one wishes to resolve in detail the post-Newtonian orbital motions of the individual SMBHs, and calculate the resulting GW spectrum \citep{2019ApJ...887...35M}.

The scalability of the \ketju{} code depends strongly on the type of problem and in particular on the relative computational cost of the regularised integration
compared to the rest of the system. For systems with a relatively low number of particles in the regularised region, the total computational cost is set by the
\gadgetfour{} part, and at high CPU counts the code might even slow down due to suboptimal scaling of the regularised integration. On the contrary, for systems with
a high particle load in the region the regularised integrator can scale nearly ideally to a much higher number of CPUs. As the total computational cost is
now dominated by the regularised integrator this allows for the use of larger CPU counts than would be sensible for a standard \gadgetfour{} simulation. In order to
avoid a slow-down of the integration at excessive CPU counts, the code automatically limits the number of CPUs
allocated to each region based on its particle count.

As a demonstration of the type of problems that can be studied using \ketju{} we run a set of simulations with multiple galaxy mergers in order
to study the formation of cores through core scouring by SMBH binaries \citep[e.g.][]{2001ApJ...563...34M,2006ApJ...648..976M,2018ApJ...864..113R, 2021ApJ...922...40D}.
In the simulations, the multiple SMBHs underwent complex dynamical interactions, with a wide range of
resulting merger timescales, determined primarily by the final eccentricities of the SMBH binaries, as expected for GW driven mergers.

All merger remnants displayed depleted stellar cores with tangentially
biased central stellar velocity dispersions and core sizes that increased with increasing numbers of progenitor galaxies.
The galaxy models were run with rather shallow Hernquist initial density profiles $(\rho \propto r^{-1})$ resulting in core sizes that
were larger than the observed cores for similar SMBH masses \citep{2016Natur.532..340T,2021ApJ...908..134D}. However, the simulations were
in good agreement with earlier simulations using similar initial conditions \citep{2018ApJ...864..113R}.
The effect of multiple successive SMBH mergers on the sizes of galactic cores has thus far not been extensively studied,
and extending the presented example study to a more complete study with more realistic initial
galaxies, multiple merger orbits with varying galactic mass ratios, and the inclusion of SMBH spins and GW recoil kicks
would be an obvious interesting application of the presented code (see e.g.\ \citealp{2021MNRAS.502.4794N,2021ApJ...922...40D} for recent simulation studies).

The public version of \gadgetfour{}, and thus by extension \ketju{}, does not include models for gas accretion onto SMBHs and the associated feedback,
or models for seeding SMBHs in cosmological simulations. Without such additional models, the presented code is mainly directly applicable for studying
SMBH dynamics in gas-poor systems, such as in the mergers of massive elliptical galaxies. For such applications the \ketju{} code allows running simulations
either at higher particle resolution or alternatively enables running a larger set of simulations than has been possible using more conventional
high-accuracy $N$-body codes
\citep[e.g.][]{2001ApJ...563...34M,2006ApJ...642L..21B,2017MNRAS.464.2301G,2021ApJ...922...40D}.

However, a variety of sophisticated models for gas cooling, star formation and evolution, as well as gas accretion onto SMBHs and the corresponding feedback
exist in the literature and have been implemented in various simulation codes.
Adding such models to the public version of \ketju{} would then allow the code to be used for a wide variety of applications, including studying SMBHs in gas-rich environments or in cosmological simulations with dynamically formed galaxies.

Alternatively, it is possible use the present code as a basis for adding the \ketju{} SMBH dynamics to other simulation codes, such as for example
AREPO \citep{2010MNRAS.401..791S}, RAMSES \citep{2002A&A...385..337T} or ENZO \citep{2014ApJS..211...19B}, which already include sophisticated subgrid
models for gas, stellar and BH physics. The effort required will depend on how similar the code architecture is to \gadgetfour{}. We also note that the existing SMBH accretion and feedback models need to be improved \citep{2023MNRAS.520.4463L},
as at $r\lesssim 100 \,\mathrm{pc}$ scale separations, where the SMBH binaries can be resolved in KETJU simulations, the preferential accretion from the circumbinary disc is important and will affect the SMBH mass ratio and the strength of the AGN feedback.

While \ketju{} is not a complete solution to modelling all aspects of how SMBH grow and interact,
we have found in our previous studies the code to be a powerful tool for studying SMBHs in certain regimes.
Future development of \ketju{}, or possibly other implementations of similar ideas, will undoubtedly further expand the regimes where
it is possible to simultaneously capture the large galactic scales and the small-scale dynamics of SMBHs.
We hope that this public release of \ketju{} will prove useful to the galactic and SMBH simulation communities
and result in new advances of our understanding of how SMBHs and galaxies interact.

\section*{Acknowledgements}
MM, AR, PHJ, DI and SL acknowledge the support by the European Research Council via ERC Consolidator Grant KETJU (no. 818930) and the support of the Academy of Finland grant 339127. AR also thanks the support of the University of Helsinki Research Foundation. TN acknowledges support from the Deutsche Forschungsgemeinschaft (DFG, German Research Foundation)
under Germany's Excellence Strategy - EXC-2094 - 390783311 from the DFG Cluster of Excellence ``ORIGIN''.
The numerical simulations used computational resources provided by
the CSC -- IT Center for Science, Finland.

\section*{Data Availability}

The code presented in this paper is available from \codewebsite{}.
The data from the presented code tests is available upon reasonable request to Peter Johansson.


\bibliographystyle{mnras}
\bibliography{refs}





\bsp	
\label{lastpage}
\end{document}